\documentclass[twocolumn,showpacs,preprintnumbers,amsmath,amssymb,aps,pre]{revtex4}

\usepackage{graphicx,color}

\begin{document}

\preprint{}

\title{Relaxation equations for two-dimensional turbulent flows \\
with a prior vorticity distribution}

 \author{P.H. Chavanis, A. Naso and B. Dubrulle}

\affiliation{$^1$ Laboratoire de Physique Th\'eorique (IRSAMC), CNRS and UPS, Universit\'e de Toulouse, F-31062 Toulouse, France\\
$^2$ Laboratoire de Physique, Ecole Normale Sup\'erieure de Lyon and CNRS (UMR 5672), 46 all\'ee d'Italie, 69007 Lyon, France\\
$^3$ SPEC/IRAMIS/CEA Saclay, and CNRS (URA 2464), 91191 Gif-sur-Yvette Cedex, France\\
\email{chavanis@irsamc.ups-tlse.fr, aurore.naso@ens-lyon.fr, berengere.dubrulle@cea.fr}
}

   \date{To be included later }

   \begin{abstract}Using a Maximum Entropy Production Principle (MEPP), we
   derive a new type of relaxation equations for two-dimensional
   turbulent flows in the case where a prior vorticity distribution is
   prescribed instead of the Casimir constraints [Ellis, Haven,
   Turkington, Nonlin., {\bf 15}, 239 (2002)]. The particular case of
   a Gaussian prior is specifically treated in connection to minimum
   enstrophy states and Fofonoff flows. These relaxation equations are
   compared with other relaxation equations proposed by Robert \&
   Sommeria [Phys. Rev. Lett. {\bf 69}, 2776 (1992)] and Chavanis
   [Physica D, {\bf 237}, 1998 (2008)]. They can serve as numerical
   algorithms to compute maximum entropy states and
   minimum enstrophy states with appropriate constraints. We perform
   numerical simulations of these relaxation equations in order to
   illustrate geometry induced phase transitions in geophysical flows.
\end{abstract}

   \maketitle
%

\section{Introduction}
\label{sec_intro}

Two-dimensional incompressible and inviscid flows are described by the 2D Euler equations
\begin{equation}
{\partial \omega\over\partial t}+{\bf u}\cdot \nabla \omega=0,\qquad \omega=-\Delta\psi,\qquad {\bf u}=-{\bf z}\times\nabla\psi,
\label{intro1}
\end{equation}
where $\omega {\bf z}=\nabla\times{\bf u}$ is the vorticity, $\psi$ the stream function and ${\bf u}$ the velocity field (${\bf z}$ is a unit vector normal to the flow).  The 2D Euler equations are known to develop a complicated mixing process which ultimately leads to the emergence of large-scale coherent structures like jets and vortices \cite{tabeling,sommeria,houches,cvh}.  The jovian atmosphere shows a wide diversity of coherent structures \cite{marcus,flierl,turk,bs,sw}: Jupiter's great red spot, white ovals, brown barges, zonal jets,... Similarly, in the earth atmosphere and in the oceans, there exists large-scale vortices such as modons (pairs of cyclones/anticyclones) or currents like the Gulf Stream or the Kuroshio Current. One question of fundamental interest is to understand and predict the structure and the stability of these quasi stationary states (QSSs). This can be done by using elements of statistical mechanics
adapted to the 2D Euler equation.

In recent years, two  statistical  theories of 2D turbulent flows have  been proposed by Miller-Robert-Sommeria (MRS) and Ellis-Haven-Turkington (EHT). These theories mainly differ in the type of constraints to be considered. Miller \cite{miller} and Robert \& Sommeria \cite{rs} assume a purely conservative evolution (no forcing and no dissipation) and take into account  all the constraints of the 2D Euler equations. The equilibrium state is obtained by maximizing a mixing entropy $S[\rho]$ while conserving energy, circulation and {\it all} the Casimirs. On the other hand, Ellis, Haven \& Turkington \cite{ellis} argue that, in real flows undergoing a permanent forcing and dissipation, some constraints are destroyed. They propose a pragmatic approach where only the robust constraints (energy and circulation) are taken into account while the fragile constraints (Casimirs) are treated canonically. This is equivalent to introducing a prior vorticity distribution $\chi(\sigma)$. This prior vorticity distribution is assumed to be determined by the properties of forcing and dissipation. The equilibrium state is then obtained by maximizing a relative entropy  $S_{\chi}[\rho]$ (depending on the prior) while conserving only energy and  circulation (robust constraints).

Some relaxation equations towards the statistical equilibrium state have been proposed in each case. Considering the MRS approach, Robert \& Sommeria \cite{rsmepp} obtained a relaxation equation   from a Maximum Entropy Production Principle (MEPP) by maximizing the production  of entropy  $S[\rho]$  at fixed energy, circulation and Casimirs. On the other hand, considering the EHT approach, Chavanis \cite{gen,physicaD,aussois} showed that the prior determines a generalized entropy $S[\overline{\omega}]$ and that the mean flow is a maximum  of generalized entropy at fixed energy and circulation. He then obtained a relaxation equation from a  MEPP by maximizing the production of  generalized  entropy  $S[\overline{\omega}]$ at fixed energy and circulation. Interestingly, the resulting equation has the form of  a nonlinear mean field Fokker-Planck (NFP) equation that appears in other domains of physics \cite{frank,nfp}.

In this paper, we introduce a new class of relaxation equations
associated with the EHT approach by maximizing the production of
relative entropy $S_{\chi}[\rho]$ at fixed energy and circulation in
order to obtain the evolution of the full distribution of vorticity
levels. Interestingly, this leads to a new class of relaxation
equations that does not appear to have been studied so far. We derive
the corresponding hierarchy of moment equations and show that it is
closed in the case of a Gaussian prior leading to a minimum enstrophy
state. These relaxation equations can serve as numerical algorithms to
compute maximum entropy states with appropriate constraints.  In the
present paper, we develop the theory and discuss in detail the link
between the MRS and the EHT approaches. We also give a numerical
illustration of our relaxation equations in relation to minimum
enstrophy states and Fofonoff flows.

The paper is organized as follows. In Secs. \ref{sec_vr},
\ref{sec_gmrs} and \ref{sec_parc} we provide a short review and
comparison of the different relaxation equations introduced in the
context of 2D turbulence in relation to the MRS and EHT statistical
theories. This review can be useful to people interested in this
topic and should clarify the subtle connections between the different
equations. We also size this opportunity to improve the discussion and
the presentation of these relaxation equations and explicitly treat
particular cases. In Sec. \ref{sec_pn}, we introduce a new class of
relaxation equations adapted to the EHT approach based on the
specification of a prior vorticity distribution. In Sec. \ref{sec_gp},
we consider the case of a Gaussian prior where these equations can be
simplified. In Sec. \ref{sec_num}, we use these relaxation equations
to illustrate phase transitions in geophysical flows, in particular the
transitions between monopoles and dipoles when the domain becomes
sufficiently stretched
\cite{jfm1,vb,Fof}.

{\it Note:} in this paper, we shall mainly follow the presentation of Ellis {\it et al.} \cite{ellis} and Chavanis \cite{gen,physicaD,aussois} who first introduced the maximization problems (\ref{gmrs3}) and (\ref{gmrs12}) based on the notion of priors. However, in Sec. \ref{sec_other}, we note that these maximixation problems also provide {\it sufficient} conditions of MRS thermodynamical stability (this is the presentation adopted by Bouchet \cite{bouchet} and Chavanis \cite{proc}). Therefore, the study of the maximization problems  (\ref{gmrs3}) and (\ref{gmrs12}), and the corresponding relaxation equations, is interesting in these {\it two} perspectives.

\section{Statistical mechanics of violent relaxation}
\label{sec_vr}

\subsection{The Miller-Robert-Sommeria theory}
\label{sec_mrs}

Starting from a generically unsteady or unstable initial condition,
the 2D Euler equation develops an intricate filamentation leading, on
the coarse-grained scale, to a quasi stationary state (QSS). The
problem is to predict the structure of this QSS as a function of the
initial condition.  A statistical theory of the 2D Euler equation has
been proposed by Miller \cite{miller} and Robert \& Sommeria \cite{rs}
(see also Kuz'min \cite{kuzmin}) by extending the approach of Onsager
based on point vortices \cite{onsager,jm,pl} or the approach of
Kraichnan based on the truncated Euler equations
\cite{kraichnan,salmon}. The Miller-Robert-Sommeria (MRS) theory is
analogous to the theory of violent relaxation developed by Lynden-Bell
\cite{lb} in the case of collisionless stellar systems described by the
Vlasov-Poisson system (see Chavanis \cite{csr,houches} for a description of the
numerous analogies between 2D vortices and stellar systems). The key
idea is to replace the deterministic description of the flow
$\omega({\bf r},t)$ by a probabilistic description where $\rho({\bf
r},\sigma,t)$ gives the density probability of finding the vorticity
level $\omega=\sigma$ in ${\bf r}$ at time $t$. The observed
(coarse-grained) vorticity field is then expressed as
$\overline{\omega}({\bf r},t)=\int \rho\sigma d\sigma$. The MRS
approach is well-suited to isolated systems (no forcing and no
dissipation) where all the inviscid invariants of the 2D Euler
equation are conserved.  The MRS statistical equilibrium state is
obtained by maximizing a mixing entropy
\begin{equation}
\label{mrs1}
S[\rho]=-\int
\rho({\bf r},\sigma)\ln \rho({\bf r},\sigma) \, d{\bf r}d\sigma,
\end{equation}
while respecting the normalization condition $\int\rho \, d\sigma=1$ and
conserving all the inviscid invariants which are the energy
\begin{equation}
\label{ene}
E={1\over 2}\int
\overline{\omega}\psi d{\bf r}={1\over 2}\int
\psi\rho\sigma \, d{\bf r}d\sigma,
\end{equation}
the circulation
\begin{equation}
\Gamma=\int
\overline{\omega}d{\bf r}=\int\rho\sigma\, d{\bf r}d\sigma,\label{circ}
\end{equation}
and the Casimirs invariants   $I_{f}=\int
\overline{f(\omega)}d{\bf r}$. This includes in particular the conservation of all the microscopic moments
of the vorticity
\begin{equation}
\Gamma_{n>1}^{f.g.}=\int
\overline{\omega^{n}}d{\bf r}=\int \rho\sigma^{n}\, d{\bf r}d\sigma.\label{mom}
\end{equation}
The conservation of the Casimirs is also equivalent to the conservation of
the fine-grained vorticity distribution $\gamma(\sigma)=\int \rho({\bf
r},\sigma)\, d{\bf r}$, i.e. the total area $\gamma(\sigma)d\sigma$
occupied by each vorticity level $\sigma$.

We shall distinguish between robust and fragile constraints. This distinction will be very important
in the following. The circulation and the
energy are called robust constraints because they can be expressed in
terms of the coarse-grained vorticity: $\overline{\Gamma[\omega]}\simeq
\Gamma[\overline{\omega}]$
and $\overline{E[\omega]}\simeq E[\overline{\omega}]$ (the energy of
the fluctuations can be neglected \cite{miller,rs}). By contrast, the
higher moments of the vorticity are called fragile constraints
because, when calculated with the coarse-grained vorticity
$\overline{\omega}$, they are not conserved since
$\overline{\omega^n}\neq \overline{\omega}^n$. Thus
$\overline{\Gamma_{n>1}[\omega]}\neq
\Gamma_{n>1}[\overline{\omega}]$. They must be expressed therefore in
terms of the ``microscopic" vorticity distribution $\rho({\bf
r},\sigma)$ as in Eq. (\ref{mom}) where we have introduced the local moments $\overline{\omega^n}=\int\rho\sigma^n\, d\sigma$ of the vorticity distribution. We must therefore distinguish the microscopic moments of the vorticity
$\Gamma_{n>1}^{f.g.}[\rho]=\int
\overline{\omega^{n}}d{\bf r}=\int \rho\sigma^{n}\, d{\bf
r}d\sigma$ (conserved) from the macroscopic moments of the vorticity
$\Gamma_{n>1}^{c.g.}[\overline{\omega}]=\int
\overline{\omega}^{n}d{\bf r}$ (non-conserved).

In the MRS approach which takes into account all the constraints of the 2D Euler equation, the
statistical equilibrium state is determined by the maximization
problem \cite{miller,rs}:
\begin{eqnarray}
\label{mrs2}
\max_{\rho}\quad \lbrace S[\rho]\quad |\quad \Gamma[\overline{\omega}]=\Gamma, \quad E[\overline{\omega}]=E, \nonumber\\
\Gamma_{n>1}^{f.g.}[\rho]=\Gamma_{n>1}^{f.g.}, \quad \int \rho \, d\sigma=1\rbrace.
\end{eqnarray}
The critical points of mixing entropy at fixed $E$, $\Gamma$, $\Gamma_{n>1}^{f.g.}$ \footnote{In principle, we should take into account the conservation of all the Casimirs, not only the moments of the vorticity. This includes integrals like $\int \overline{|\omega|}\, d{\bf r}$. This extension is straightforward but for ease of notations we shall only consider the vorticity moments.} and normalization are obtained from the variational principle
\begin{equation}
\label{mrs3}
\delta
S-\beta\delta E-\alpha\delta\Gamma-\sum_{n>1}\alpha_{n}\delta\Gamma_{n}^{f.g.}-\int
\zeta({\bf r})\delta\rho \, d\sigma d{\bf r}=0,
\end{equation}
where $\beta$, $\alpha$, $\alpha_{n>1}$ and $\zeta({\bf r})$ are appropriate Lagrange
multipliers. This leads to the Gibbs state
\begin{equation}
\label{mrs4}
\rho({\bf r},\sigma)={1\over Z({\bf r})}\chi(\sigma)  e^{-(\beta\psi+\alpha)\sigma},
\end{equation}
where $ Z=\int \chi(\sigma)
e^{-(\beta\psi+\alpha)\sigma}d\sigma$ is a normalization factor and we
have defined the function $\chi(\sigma)\equiv{\rm
exp}(-\sum_{n>1}\alpha_{n}\sigma^n)$ which encapsulates the Lagrange
multipliers associated with the fragile constraints. The
coarse-grained vorticity is then given by
\begin{eqnarray}
\label{mrs5}
\overline{\omega}={\int \chi(\sigma)\sigma e^{-(\beta\psi+\alpha)\sigma} d\sigma\over \int \chi(\sigma) e^{-(\beta\psi+\alpha)\sigma} d\sigma}=-\frac{1}{\beta}\frac{d\ln Z}{d\psi}=F(\beta\psi+\alpha).\
\end{eqnarray}
The function $F$ is
explicitly given by
\begin{eqnarray}
\label{deff}
F(\Phi)=-(\ln\hat\chi)'(\Phi),
\end{eqnarray}
where we have
defined $\hat{\chi}(\Phi)\equiv\int\chi(\sigma) e^{-\sigma \Phi}d\sigma$.
On the other hand, differentiating Eq. (\ref{mrs5}) with respect to $\psi$, it is easy to show that the local centered variance
of the vorticity distribution
\begin{equation}
\label{locvar}
\omega_2\equiv \overline{\omega^2}-\overline{\omega}^2=\int \rho(\sigma-\overline{\omega})^2\, d{\bf r},
\end{equation}
is given by
\cite{sw,physicaD}:
\begin{equation}
\label{mrs6}
\omega_{2}=-\frac{1}{\beta}\overline{\omega}'(\psi)=\frac{1}{\beta^2}\frac{d^2\ln Z}{d\psi^2}.
\end{equation}
As noted in \cite{sw}, this relation bears some formal similarities with the fluctuation-dissipation theorem (FDT). Since $\overline{\omega}=\overline{\omega}(\psi)$, the statistical
theory predicts that the coarse-grained vorticity
$\overline{\omega}({\bf r})$ is a {stationary solution} of the 2D
Euler equation. On the other hand, since
$\overline{\omega}'(\psi)=-\beta\omega_{2}(\psi)$ with $\omega_2\ge
0$, the $\overline{\omega}-\psi$ relationship is a {monotonic}
function that is increasing at {negative temperatures} $\beta<0$ and
decreasing at positive temperatures $\beta>0$. Therefore, the
statistical theory {predicts} that the QSS is characterized by
a monotonic $\overline{\omega}(\psi)$ relationship. This
$\overline{\omega}-\psi$ relationship can take different shapes
depending on the initial condition. Substituting Eq. (\ref{mrs5}) in
the Poisson equation (\ref{intro1}), the equilibrium state is obtained
by solving the differential equation
\begin{equation}
\label{mrs7}
-\Delta\psi=-\frac{1}{\beta}\frac{d}{d\psi}\ln Z,
\end{equation}
with $\psi=0$ on the boundary of the domain and relating the Lagrange multipliers to
the constraints. Then, we have to make sure that the distribution (\ref{mrs4}) is a (local) maximum
of entropy, not a minimum or a saddle point. A critical point of constrained entropy
is a (local) maximum iff \cite{proc}:
\begin{eqnarray}
\delta^{2}J\equiv -\frac{1}{2}\int \frac{(\delta \rho)^{2}}{\rho} \, d{\bf r}d\sigma-\frac{\beta}{2}\int \delta\overline{\omega}\delta\psi  \, d{\bf r}<0,\nonumber\\
\forall \delta\rho\ | \ \delta E=\delta\Gamma=\delta\Gamma_{n>1}^{f.g.}=\int \delta\rho\, d\sigma=0,\qquad
\label{ts8}
\end{eqnarray}
i.e. for all perturbations $\delta\rho$ that conserve the constraints (circulation, energy, Casimirs, normalization) at first order. Finally, if several (local) entropy maxima remain for the same values of the constraints, we can compare their entropies to determine which one is the global entropy maximum and which one is a relative entropy maximum.
We stress, however, that local entropy maxima can be long-lived, hence fully relevant, for systems with long-range interactions\footnote{In fact, even saddle points of entropy may be relevant because it can take time for the system to find the optimal perturbation(s) that destabilizes them \cite{minens}.}.

\subsection{Relaxation equations}
\label{sec_relax}

Robert \& Sommeria \cite{rsmepp} have introduced a relaxation
equation solving the optimization problem (\ref{mrs2}) by maximizing
the rate of entropy production $\dot S$ at fixed circulation, energy
and Casimir constraints (and other physical constraints putting
a  bound on the diffusion currents). This Maximum
Entropy Production Principle (MEPP) leads to the following relaxation
equation
\begin{equation}
\label{mrs8}
\frac{\partial\rho}{\partial t}+{\bf u}\cdot\nabla\rho=\nabla\cdot\left\lbrack D\left (\nabla\rho+\beta(t)(\sigma-\overline{\omega})\rho\nabla\psi\right )\right\rbrack,
\end{equation}
where
\begin{equation}
\beta(t)=-{\int D\nabla\overline{\omega}\cdot\nabla\psi \, d{\bf r}\over \int D\omega_2(\nabla\psi)^{2}\, d{\bf r}},
\label{mrs9}
\end{equation}
is a Lagrange multiplier (inverse temperature) enforcing the energy constraint $\dot E=0$ at
any time and $D({\bf r},t)$ is a diffusion coefficient. The diffusion coefficient is not determined by the MEPP but it must be positive to have an increase of entropy (see below). The boundary conditions are ${\bf
J}\cdot {\bf n}=0$ where ${\bf J}=-D
(\nabla\rho+\beta(t)(\sigma-\overline{\omega})\rho\nabla\psi)$
is the current of level $\sigma$ and ${\bf n}$ is a unit vector normal
to the boundary. Easy calculations lead to the H-theorem
\begin{equation}
\dot S=\int \frac{D}{\rho}\left (\nabla\rho+\beta(t)\rho (\sigma-\overline{\omega})\nabla\psi\right )^{2}\, d{\bf r}d\sigma\ge 0.
\label{mrs10}
\end{equation}
Equation (\ref{mrs8}) with the constraint (\ref{mrs9}) has the
following properties: (i) $\Gamma$, $E$, $\Gamma_{n>1}^{f.g.}$ and
$\int \rho d\sigma=1$ are conserved. (ii) $\dot S\ge 0$. (iii) $\dot
S=0$ $\Leftrightarrow$ $\rho({\bf r},\sigma)$ is the Gibbs state
(\ref{mrs4}) $\Leftrightarrow$ $\partial_{t}\rho=0$. (iv) $\rho({\bf
r},\sigma)$ is a steady state of Eqs. (\ref{mrs8})-(\ref{mrs9}) iff it
is a critical point of $S$ at fixed $E$, $\Gamma$,
$\Gamma_{n>1}^{f.g.}$ and normalization. (v) A steady state of
Eqs. (\ref{mrs8})-(\ref{mrs9}) is linearly stable iff it is a (local)
maximum of $S$ at fixed $E$, $\Gamma$, $\Gamma_{n>1}^{f.g.}$ and
normalization. By Lyapunov's direct method, we know that if $S$ is
bounded from above, Eqs. (\ref{mrs8})-(\ref{mrs9}) will relax towards
a (local) maximum of $S$ at fixed $E$, $\Gamma$, $\Gamma_{n>1}^{f.g.}$
and normalization (if several local maxima exist, the choice of the
maximum will depend on a notion of basin of attraction). Therefore, a
stable steady state of Eqs. (\ref{mrs8})-(\ref{mrs9}) solves the
maximization problem (\ref{mrs2}). By construction, it is the MRS
statistical equilibrium state (most mixed state) corresponding to a
given initial condition. As a result, the relaxation equations
(\ref{mrs8})-(\ref{mrs9}) can  serve as a numerical algorithm to solve the
maximization problem (\ref{mrs2}) for a given value of the constraints
specified by the initial condition. These relaxation equations have been
studied theoretically and numerically in \cite{rr,rm,ros1,ros2}.

{\it Remark 1:} the relaxation equations (\ref{mrs8})-(\ref{mrs9}) do
not respect the {\it invariance properties} of the 2D Euler equation
(invariance by translation or rotation) and
this may be a serious problem to describe the evolution of the flow
into several isolated vortices (if we use these equations as a
parametrization of 2D turbulence). A solution to this problem has been
proposed by Chavanis
\& Sommeria \cite{csmepp} by reformulating the MEPP under a local
form, introducing currents of energy, angular momentum and
impulse. However, the resulting relaxation equations are more
complicated and have not been numerically solved for the moment.

{\it Remark 2:} in order to take into account incomplete relaxation \cite{jfm2,brands}, Robert \& Rosier \cite{rr} and Chavanis {\it et al.} \cite{csr} have proposed to use a diffusion coefficient depending on the local fluctuations of vorticity. This can freeze the system in a ``maximum entropy bubble'' which is a restricted maximum entropy state \cite{jfm2}. This diffusion coefficient can   also be justified from a quasilinear theory of the 2D Euler equation \cite{quasi}. It can be written in the form \cite{rr,csr}:
\begin{equation}
D({\bf r},t)=K\epsilon^2\omega_2^{1/2},
\label{mrs11d}
\end{equation}
where $K$ is a constant of order unity, $\epsilon$ the scale of unresolved fluctuations and $\omega_2$ the local centered variance of the vorticity.

\subsection{Moment equations}
\label{sec_moment}

From Eq. (\ref{mrs8}), we can derive a hierarchy of
equations for the local moments of the vorticity
$\overline{\omega^{n}}=\int \rho\sigma^{n}\, d\sigma$. The equation for the moment of order $n$ is
\begin{eqnarray}
{\partial \overline{\omega^n}\over\partial t}&+&{\bf u}\cdot \nabla
\overline{\omega^n}\nonumber\\
&=&\nabla\cdot \biggl \lbrace D
\biggl\lbrack \nabla\overline{\omega^n}+\beta(t)(\overline{\omega^{n+1}}
-\overline{\omega}\overline{\omega^n})\nabla\psi\biggr\rbrack\biggr\rbrace.\qquad
\end{eqnarray}
This hierarchy of equations is not closed since the equation for the moment of order $n$ involves the moment of order $n+1$. Robert \& Rosier \cite{rr} have proposed to close the hierarchy of equations by assuming that the density distribution $\rho({\bf r},\sigma,t)$ maximizes the entropy (\ref{mrs1}) with the constraints of the known first $n$ moments and the normalization. This gives a density of the form $\rho\sim {\rm exp}(-\sum_{k=1}^{n}\alpha_k\sigma^{k})$, where the Lagrange multipliers $\alpha_k$ can be calculated from the constraints of the known moments. Then, $\overline{\omega^{n+1}}$ can be obtained from this distribution and expressed in terms of $\overline{\omega}$,...,$\overline{\omega^n}$.

Kazantsev {\it et al.} \cite{kazantsev} have considered in detail the case $n=2$. If we maximize the entropy (\ref{mrs1}) at fixed  $\overline{\omega}({\bf r},t)=\int \rho\sigma\, d\sigma$ and $\overline{\omega^2}({\bf r},t)=\int \rho\sigma^2\, d\sigma$, we obtain a Gaussian distribution of the form
\begin{equation}
\rho({\bf r},\sigma,t)=\frac{1}{\sqrt{2\pi\omega_2}}e^{-\frac{(\sigma-\overline{\omega})^2}{2\omega_2}}.\label{gauss}
\end{equation}
Therefore, the vorticity distribution  is locally Gaussian with mean value  $\overline{\omega}({\bf r},t)$ and centered variance $\omega_2({\bf r},t)$. From this distribution, we compute $\overline{\omega^3}=\overline{\omega}^3+3\overline{\omega}\omega_2$ which closes the hierarchy at the order $n=2$. Then, the equations for the mean vorticity and the centered variance can be conveniently written
\begin{equation}
{\partial \overline{\omega}\over\partial t}+{\bf u}\cdot \nabla
\overline{\omega}=\nabla\cdot \biggl \lbrace D
\biggl\lbrack \nabla\overline{\omega}+\beta(t)\omega_2\nabla\psi\biggr\rbrack\biggr\rbrace.
\label{mrs11}
\end{equation}
\begin{eqnarray}
{\partial \omega_2\over\partial t}+{\bf u}\cdot \nabla
\omega_2&=&\nabla\cdot (D\nabla\omega_2)\nonumber\\
&+&2D\nabla\overline{\omega}\cdot\left ( \nabla\overline{\omega}+\beta(t)\omega_2\nabla\psi\right ).\label{om2}
\end{eqnarray}
These equations conserve by construction the energy, the circulation and the microscopic enstrophy $\Gamma_2^{f.g.}=\int\overline{\omega^2}\, d{\bf r}$ but not the higher moments. On the other hand, we can prove an $H$-theorem for the entropy. Using the Gaussian distribution (\ref{gauss}), the mixing entropy (\ref{mrs1}) can be written
\begin{eqnarray}
\label{entom}
S=\frac{1}{2}\int \ln\omega_2\, d{\bf r},
\end{eqnarray}
up to some unimportant additive constants. Then, using Eqs. (\ref{mrs11}) and (\ref{om2}),  it is easy to establish the H-theorem
\begin{eqnarray}
\dot S=\int \frac{D}{2\omega_2^2}(\nabla\omega_2)^2\, d{\bf r}+\int \frac{D}{\omega_2}(\nabla\overline{\omega}+\beta(t)\omega_2\nabla\psi)^2\, d{\bf r}\ge 0.\nonumber\\
\end{eqnarray}
At equilibrium, $\dot S=0$, we obtain
\begin{eqnarray}
\nabla\overline{\omega}+\beta\omega_2\nabla\psi={\bf 0},
\end{eqnarray}
\begin{eqnarray}
\nabla\omega_2={\bf 0}.
\end{eqnarray}
The second equation shows that the centered variance $\omega_2({\bf r})=\Omega_2$ is uniform and the first equation can then be integrated into
\begin{eqnarray}
\overline{\omega}=-\Omega_2(\beta\psi+\alpha),
\label{ajy}
\end{eqnarray}
where $\alpha$ is a constant.
The statistical equilibrium state presents a mean flow characterized by a linear $\overline{\omega}-\psi$ relationship and Gaussian fluctuations around it. It is a maximum of entropy at fixed energy $E$, circulation $\Gamma$ and microscopic enstrophy $\Gamma_2^{f.g.}$. It is also a minimum of macroscopic  enstrophy $\Gamma_2^{c.g.}$ at fixed energy $E$ and circulation $\Gamma$  (see Appendix \ref{sec_meh}).

{\it Remark:} We note that, with this formalism, it is technically difficult to go beyond the Gaussian closure approximation. The approach developed in Sec. \ref{sec_parc} may provide an alternative strategy to describe more complex situations where the vorticity distribution is not Gaussian.

\subsection{The equation for the velocity field}
\label{sec_nept}

Following Chavanis \& Sommeria \cite{sw}, we can derive a relaxation equation for the velocity field. The equation for the coarse-grained vorticity field is
\begin{equation}
{\partial \overline{\omega}\over\partial t}+{\bf u}\cdot \nabla
\overline{\omega}=\nabla\cdot \biggl \lbrace D
\biggl\lbrack \nabla\overline{\omega}+\beta(t)\omega_2\nabla\psi\biggr\rbrack\biggr\rbrace,
\label{mrs11bis}
\end{equation}
where we recall that $D({\bf r},t)$ can depend on position and time. For a 2D field, we have the identity  $\nabla\times ({\bf z}\times {\bf a})=(\nabla\cdot {\bf a}){\bf z}$. Therefore, we can rewrite the foregoing equation as
\begin{eqnarray}
\label{nept2}
\left (\frac{\partial \overline{\omega}}{\partial t}+{\bf u}\cdot \nabla \overline{\omega}\right ) {\bf z}=\nabla\times  \left\lbrack {\bf z}\times  D\left (\nabla \overline{\omega}+{\beta(t)}\omega_2\nabla\psi\right )\right\rbrack. \nonumber\\
\end{eqnarray}
Since $\nabla\times {\bf u}=\overline{\omega} {\bf z}$, the corresponding equation for the velocity field is
\begin{eqnarray}
\label{nept3}
\frac{\partial {\bf u}}{\partial t}+({\bf u}\cdot \nabla) {\bf u}=-\frac{1}{\rho}\nabla p+ {\bf z}\times  D\left (\nabla \overline{\omega}+\beta(t)\omega_2\nabla\psi\right ),\nonumber\\
\end{eqnarray}
where $p$ is the pressure and $\rho$ the density. Now, using ${\bf u}=-{\bf z}\times \nabla\psi$ and the identity
\begin{eqnarray}
\label{nept4}
\Delta {\bf u}=\nabla (\nabla\cdot {\bf u})-\nabla\times(\nabla\times {\bf u})\nonumber\\
=-\nabla\times(\overline{\omega} {\bf z})={\bf z}\times \nabla\overline{\omega},
\end{eqnarray}
valid for a 2D incompressible flow, we finally obtain
\begin{eqnarray}
\label{nept5}
\frac{\partial {\bf u}}{\partial t}+({\bf u}\cdot \nabla) {\bf u}=-\frac{1}{\rho}\nabla p+   D\left (\Delta {\bf u}-{\beta(t)}\omega_2{\bf u}\right ).
\end{eqnarray}
We see that the drift term in the equation for the vorticity (\ref{mrs11bis}) takes the form of a friction in the equation for the velocity (\ref{nept5}). Furthermore, the drift coefficient or the friction coefficient is given by an Einstein-like formula $\xi=D\beta$ involving the diffusion coefficient and the inverse temperature. At equilibrium, we get
\begin{eqnarray}
\label{nept6}
\Delta {\bf u}={\beta}\omega_2{\bf u},
\end{eqnarray}
which can be directly derived from the Gibbs state (\ref{mrs4}) using $\Delta {\bf u}={\bf z}\times \nabla\overline{\omega}$, $\nabla\overline{\omega}=\overline{\omega}'(\psi)\nabla\psi$ and Eq. (\ref{mrs6}).

\section{Statistical mechanics with a prior vorticity distribution}
\label{sec_gmrs}

\subsection{The Ellis-Haven-Turkington approach}
\label{sec_gmrseht}

In the MRS theory, it is assumed that the flow is rigorously described
by the 2D Euler equation so that all the Casimirs are
conserved. However, in many geophysical situations, the flows are
forced and dissipated at small scales (e.g., due to convection in the
jovian atmosphere) so that the conservation of the fragile constraints
(Casimirs) is destroyed \cite{turk}. Ellis, Haven and Turkington
\cite{ellis} have proposed to treat these situations by keeping only
the robust constraints $E$ and $\Gamma$ and replacing the conservation
of the fragile constraints $\Gamma_{n>1}^{f.g.}$ by the specification
of a prior vorticity distribution $\chi(\sigma)$. As noted by Chavanis
\cite{physicaD}, this amounts to making a Legendre transform of the
MRS entropy (\ref{mrs1}) with respect to the fragile constraints (see
Eq. (\ref{gmrs1})). The EHT approach corresponds therefore to a grand
microcanonical version \cite{proc} of the MRS theory in which the
chemical potentials $\alpha_{n>1}$ associated with the microscopic
constraints are given\footnote{The EHT approach
implicitly assumes that the system is in contact with a ``bath''
fixing the chemical potentials instead of the Casimirs. It is not
clear, especially for systems with long-range interactions, how this
notion of bath can be rigorously defined. Furthermore, it is not clear
whether a system undergoing a permanent forcing and dissipation can be
described by equilibrium statistical mechanics since it does not, in
principle, verify a detailed balance property. It is therefore
important to determine to which physical situations this approach can
be applied. In that respect, we can mention the recent paper of
Dubinkina \& Frank
\cite{df} that shows that the EHT approach can explain successfully
some results of numerical simulations.}.

We introduce the grand entropy
\cite{physicaD}:
\begin{equation}
\label{gmrs1}
S_{\chi}[\rho]=S[\rho]-\sum_{n>1}\alpha_{n}\Gamma_{n}^{f.g.}.
\end{equation}
Explicitly, we have
\begin{equation}
\label{gmrs2}
S_{\chi}\lbrack \rho\rbrack=-\int \rho({\bf r},\sigma)\ \ln\biggl\lbrack {\rho({\bf r},\sigma)\over \chi(\sigma)}\biggr\rbrack \ d{\bf r}d\sigma,
\end{equation}
where $\chi(\sigma)={\rm exp}(-\sum_{n>1}\alpha_{n}\sigma^n)$. In the
present context, this function is {\it given} and is
called the prior vorticity distribution. It is determined by the properties of forcing and dissipation for the situation considered. On the other hand, $S_{\chi}$ is called the relative entropy \cite{turk}.
The EHT statistical equilibrium state is
obtained by maximizing the relative (or grand) entropy (\ref{gmrs2}) while
respecting the normalization condition $\int\rho d\sigma=1$ and
conserving only the {\it robust} constraints $\Gamma=\int
\overline{\omega}d{\bf r}$ and $E={1\over 2}\int
\overline{\omega}\psi d{\bf r}$. Therefore, we have to solve the maximization problem \cite{ellis}:
\begin{eqnarray}
\label{gmrs3}
\max_{\rho}\quad \lbrace S_{\chi}[\rho]\quad |\quad \Gamma[\overline{\omega}]=\Gamma, \quad E[\overline{\omega}]=E, \quad \int \rho d\sigma=1\rbrace.\nonumber\\
\end{eqnarray}

The critical points of grand entropy $S_{\chi}$ at fixed
circulation, energy and normalization (canceling the first variations) are given by the variational principle
\begin{equation}
\label{gmrs4}
\delta
S_{\chi}-\beta\delta E-\alpha\delta\Gamma-\int
\zeta({\bf r})\delta\rho d\sigma d{\bf r}=0.
\end{equation}
This leads to the Gibbs state (\ref{mrs4}). Therefore, the critical points of the variational principles  (\ref{mrs2})  and  (\ref{gmrs3}) coincide. On the other hand, a critical point of constrained grand entropy is a (local) maximum iff \cite{proc}:
\begin{eqnarray}
\delta^{2}J\equiv -\frac{1}{2}\int \frac{(\delta \rho)^{2}}{\rho} \, d{\bf r}d\sigma-\frac{\beta}{2}\int \delta\overline{\omega}\delta\psi  \, d{\bf r}<0,\nonumber\\
\forall \delta\rho\ | \ \delta E=\delta\Gamma=\int \delta\rho\, d\sigma=0,\qquad
\label{gmrs8}
\end{eqnarray}
i.e., for all perturbations $\delta\rho$ that conserve circulation, energy
and normalization at first order. This differs from (\ref{ts8}) at the level of the class of perturbations to be considered. We shall come back to the connection between the MRS and EHT theories in Sec. \ref{sec_other}.

\subsection{Generalized entropies}
\label{sec_get}

We shall now introduce a reduced variational problem equivalent to (\ref{gmrs3}) but expressed in terms of a generalized entropy $S[\overline{\omega}]$ associated with the coarse-grained
flow instead of a functional $S_{\chi}[\rho]$ associated to the full vorticity distribution. Initially, we want to determine the
vorticity distribution $\rho_*({\bf r},\sigma)$ that maximizes
$S_{\chi}[\rho]$ with the robust constraints $E[\overline{\omega}]=E$,
$\Gamma[\overline{\omega}]=\Gamma$ and the normalization condition
$\int\rho\, d\sigma=1$. To solve this maximization problem (\ref{gmrs3}), we can proceed in two steps\footnote{This ``two-steps'' approach was developed by one of us (PHC)  in different situations of fluid mechanics and astrophysics (see, e.g., \cite{assise}).}.

{\it (i) First step:} We first  determine the distribution
$\rho_1({\bf r},\sigma)$ that maximizes $S_{\chi}[\rho]$ with the
constraints $E$, $\Gamma$, $\int\rho\, d\sigma=1$ {\it and} a fixed vorticity profile
$\overline{\omega}({\bf r})=\int\rho\sigma \, d\sigma$. Since the
specification of $\overline{\omega}({\bf r})$ determines $\Gamma$ and $E$, this
is equivalent to  maximizing $S_{\chi}[\rho]$ with the
constraints $\int\rho\, d\sigma=1$ and $\int\rho\sigma \, d\sigma =\overline{\omega}({\bf r})$.
Writing the  first order variations as
\begin{eqnarray}
\delta S_{\chi}-\int \Phi({\bf
r}) \delta \left (\int \rho \sigma d\sigma\right ) d{\bf r}-\int \zeta({\bf r})
\delta \left (\int \rho d\sigma\right ) d{\bf r}=0,\nonumber\\
\end{eqnarray}
where $\Phi({\bf r})$ and
$\zeta({\bf r})$ are Lagrange multipliers, we obtain
\begin{eqnarray}
\rho_1({\bf r},\sigma)=\frac{1}{Z({\bf
r})}\chi(\sigma)e^{-\sigma\Phi({\bf r})},\label{ge1}
\end{eqnarray}
where $Z({\bf r})$ and $\Phi({\bf r})$ are determined by the contraints $\int\rho \, d\sigma=1$ and $\overline{\omega}=\int\rho\sigma\, d\sigma$ leading to
\begin{eqnarray} Z({\bf r})=\int \chi(\sigma)e^{-\sigma\Phi({\bf
r})}d\sigma\equiv \hat{\chi}(\Phi),\label{ge2}
\end{eqnarray}
\begin{eqnarray}
\overline{\omega}({\bf r})=\frac{1}{Z({\bf r})}\int \chi(\sigma)\sigma
e^{-\sigma\Phi({\bf r})}d\sigma=-(\ln\hat{\chi})'(\Phi).\label{ge3}
\end{eqnarray}
Equation (\ref{ge3}) relates $\Phi$ to  the vorticity profile  $\overline{\omega}$ and Eq. (\ref{ge2}) determines
$Z$. The critical point (\ref{ge1}) is a {\it maximum} of $S_{\chi}$ with the
above-mentioned constraints since $\delta^{2} S_{\chi}=-\int
\frac{(\delta\rho)^{2}}{2\rho} d{\bf r}d\sigma< 0$ (the constraints are linear in $\rho$ so their second variations vanish).
This gives a distribution $\rho_1[\overline{\omega}({\bf
r}),\sigma]$ depending on $\overline{\omega}({\bf r})$ and
$\sigma$. Substituting this distribution in the functional
$S_{\chi}[\rho]$, we obtain a functional $S[\overline{\omega}]\equiv
S_{\chi}[\rho_1]$ of the vorticity $\overline{\omega}$ alone.
Using Eqs. (\ref{gmrs2}) and (\ref{ge1}), it is given by
\begin{eqnarray}
S[\overline{\omega}]=\int \rho_1(\sigma\Phi+\ln\hat\chi )\, d{\bf
r}d\sigma=\int (\overline{\omega}\Phi+\ln\hat{\chi}(\Phi))  \, d{\bf
r}.\nonumber\\
\label{ge4}
\end{eqnarray}
Therefore, $S[\overline{\omega}]$ can be written
\begin{eqnarray}
\label{genent}
S[\overline{\omega}]=-\int C(\overline{\omega}) \, d{\bf r},
\end{eqnarray}
with
\begin{eqnarray}
C(\overline{\omega})= -\overline{\omega}\Phi-\ln\hat{\chi}(\Phi).\label{ge6}
\end{eqnarray}
Now, $\Phi({\bf r})$ is related to $\overline{\omega}(\bf{r})$ by Eq. (\ref{ge3}).
This implies that
\begin{eqnarray}
C'(\overline{\omega})=-
\Phi=-[(\ln\hat\chi)']^{-1}(-\overline{\omega}),\label{ge7}
\end{eqnarray}
so that
\begin{eqnarray}
C(\overline{\omega})=-\int^{\overline{\omega}}
[(\ln\hat\chi)']^{-1}(-x)dx.\label{ge8}
\end{eqnarray}
This can be written equivalently
\begin{eqnarray}
\label{gmrs11} C(\overline{\omega})=-\int^{\overline{\omega}}F^{-1}(x)dx
\end{eqnarray}
where the function $F$ is defined by Eq. (\ref{deff}). Note that the function  $C$ is convex, i.e. $C''>0$. Equation (\ref{genent}) with (\ref{ge8})  is the  entropy of the coarse-grained vorticity. It is completely specified by the prior $\chi(\sigma)$. Since the function $C$ can take several forms depending on the prior, $S$ is sometimes called a generalized entropy \cite{gen,physicaD,aussois}. This model of 2D turbulence is therefore an interesting physical example where generalized forms of entropy can arise \cite{frank,nfp}.

Before going further, let us establish some useful identities. From Eqs. (\ref{ge1})-(\ref{ge3}), we easily obtain
\begin{eqnarray}
\omega_2(\Phi)=-\overline{\omega}'(\Phi).\label{idp}
\end{eqnarray}
On the other hand, taking the derivative of Eq. (\ref{ge7}), we have
\begin{eqnarray}
C''(\overline{\omega})=-\frac{d\Phi}{d\overline{\omega}}=-\frac{1}{\overline{\omega}'(\Phi)}.\label{ge7b}
\end{eqnarray}
Combining Eqs. (\ref{idp}) and (\ref{ge7b}), we get \cite{physicaD}:
\begin{eqnarray}
\omega_2=\frac{1}{C''(\overline{\omega})}. \label{ind}
\end{eqnarray}

{\it (ii) Second step:}  we now have to determine the vorticity field $\overline{\omega}_*({\bf r})$
that maximizes $S[\overline{\omega}]$ with the constraints
$E[\overline{\omega}]=E$ and
$\Gamma[\overline{\omega}]=\Gamma$. We thus consider the maximization problem
\begin{eqnarray}
\label{gmrs12}
\max_{\overline{\omega}}\quad \lbrace S[\overline{\omega}]\quad |\quad \Gamma[\overline{\omega}]=\Gamma, \quad E[\overline{\omega}]=E \rbrace.
\end{eqnarray}
The critical points of $S[\overline{\omega}]$ at fixed $E$ and $\Gamma$
satisfy the variational principle
\begin{eqnarray}
\label{cbb}
\delta S-\beta\delta E-\alpha\delta
\Gamma=0,
\end{eqnarray}
where $\beta$ and $\alpha$ are Lagrange multipliers. This
yields
\begin{eqnarray}
C'(\overline{\omega})=-\beta\psi-\alpha.
\label{cba}
\end{eqnarray}
Using Eq. (\ref{gmrs11}) this is equivalent to
$\overline{\omega}=F(\beta\psi+\alpha)$ and we recover the
coarse-grained vorticity (\ref{mrs5}) deduced from the Gibbs state
(\ref{mrs4}). Differentiating the previous relation, we note that
\begin{eqnarray}
\label{css}
\overline{\omega}'(\psi)=-\frac{\beta}{C''(\overline{\omega})}.
\end{eqnarray}
According to Eq. (\ref{ge7}), we also note that Eq. (\ref{cba}) is equivalent to
\begin{eqnarray}
\Phi=\beta\psi+\alpha,
\label{cbab}
\end{eqnarray}
at equilibrium.
Then, the identity (\ref{idp}) becomes
\begin{eqnarray}
\omega_2=-\frac{1}{\beta}\overline{\omega}'(\psi),\label{idpb}
\end{eqnarray}
returning Eq.  (\ref{mrs6}).  Comparing Eqs. (\ref{css}) and (\ref{idpb}), we recover Eq. (\ref{ind}).
On the other hand, a critical point of (\ref{gmrs12}) is a maximum of $S$ at fixed $E$ and $\Gamma$ iff \cite{proc}:
\begin{equation}
\label{gmrs13}
\delta^2 J=-\frac{1}{2}\int C''(\overline{\omega})(\delta\overline{\omega})^2\, d{\bf r}-\frac{1}{2}\beta\int\delta\overline{\omega}\delta\psi\, d{\bf r}<0,
\end{equation}
for all variations $\delta\overline{\omega}$ that conserve circulation and energy at first order.

{\it (iii) Conclusion:} Finally, the solution $\rho_*({\bf
r},\sigma)$ of (\ref{gmrs3}) is given by Eq. (\ref{ge1}) where $\omega_*({\bf r})$ is the solution of (\ref{gmrs12}).  Therefore, $\rho_{*}({\bf
r},\sigma)=\rho_1[\overline{\omega}_*({\bf r}),\sigma]$  maximizes
$S_{\chi}[\rho]$ at fixed $E$, $\Gamma$ and normalization iff
$\overline{\omega}_{*}({\bf r})$ maximizes
$S[\overline{\omega}]$ at fixed $E$ and $\Gamma$. Therefore, (\ref{gmrs3}) and (\ref{gmrs12}) are equivalent but (\ref{gmrs12}) is simpler to study because it is expressed in terms of the vorticity field $\overline{\omega}({\bf r})$ instead of the full vorticity distribution $\rho({\bf r},\sigma)$. The equivalence between the stability conditions (\ref{gmrs8}) and (\ref{gmrs13}) is shown explicitly in Appendix \ref{sec_eqa}.

In conclusion, in the EHT approach, the statistical equilibrium state $\rho({\bf r},\sigma)$ maximizes a relative entropy $S_{\chi}[\rho]$ at fixed circulation $\Gamma$, energy $E$ and normalization condition if and only if the equilibrium coarse-grained field $\overline{\omega}({\bf r})$ maximizes a generalized entropy $S[\overline{\omega}]$ (determined by the prior) at fixed circulation $\Gamma$ and energy $E$. We have the equivalence
\begin{eqnarray}
\label{eqa}
(\ref{gmrs3}) \Leftrightarrow (\ref{gmrs12}).
\end{eqnarray}
This provides a condition of thermodynamical stability in the EHT sense.  On the other hand, Ellis-Haven-Turkington \cite{ellis} have shown that the maximization problem (\ref{gmrs12}) also provides a refined condition of nonlinear dynamical stability with respect to the 2D Euler equation (see \cite{proc} for further discussion). Therefore, a EHT statistical equilibrium state is both thermodynamically stable (with respect to variations of the fine-grained vorticity distribution $\delta\rho$) and nonlinearly dynamically stable (with respect to variations of the coarse-grained vorticity field $\delta\overline{\omega}$).

\subsection{Another interpretation of the EHT approach}
\label{sec_other}

As noted by Bouchet \cite{bouchet}, and further discussed by Chavanis \cite{proc}, there exists another interpretation of the EHT approach. As we have already indicated, the EHT approach can be interpreted as a grand microcanonical version of the MRS theory in which the Lagrange multipliers $\alpha_{n>1}$ are fixed instead of the constraints $\Gamma^{f.g.}_{n>1}$ \cite{physicaD}. Therefore, the MRS theory is associated to the microcanonical ensemble while the EHT approach is associated to the grand microcanonical ensemble \cite{proc}. Now it is well-known in statistical mechanics that a solution of a maximization problem is always solution of a more constrained dual maximization problem \cite{ellisineq}. In particular, grand
microcanonical stability implies microcanonical stability. Therefore, the EHT condition of thermodynamical stability provides  a sufficient (but not necessary) condition of MRS thermodynamical stability. We have the implication
\begin{eqnarray}
 (\ref{gmrs3}) \Rightarrow (\ref{mrs2}).
\label{gmrs9}
\end{eqnarray}
This implication can be directly obtained from the stability
conditions (\ref{ts8}) and (\ref{gmrs8}). Indeed, if inequality
(\ref{gmrs8}) is satisfied for all perturbations
that conserve circulation, energy and normalization, then it is
satisfied {\it a fortiori} for perturbations that
conserve circulation, energy, normalization and all the Casimirs, so
that (\ref{ts8}) is fulfilled.  Therefore, an EHT equilibrium is
always a MRS equilibrium but the converse is wrong because some
constraints have been treated canonically. This is related to the
notion of ensemble inequivalence in thermodynamics for systems with
long-range interactions \cite{ellisineq,bb,revue}. There can exist
states that solve the maximization problem (\ref{mrs2}) although they
do not solve (\ref{gmrs3}). Such states cannot be reached by a grand
microcanonical description. In that case, we have ensemble
inequivalence. Therefore, an interpretation of the optimization
problem (\ref{gmrs3}) is that it provides a {\it sufficient} condition
of MRS thermodynamical stability.

On the other hand,  the coarse-grained vorticity field associated to a maximum
of $S[\rho]$ at fixed circulation, energy, Casimirs and normalization
is always a critical point of  $S[\overline{\omega}]$ at fixed
circulation and energy. However, it is not necessarily a {\it maximum} of $S[\overline{\omega}]$ at fixed
circulation and energy since (\ref{gmrs12}) is not equivalent to (\ref{mrs2}). According to (\ref{eqa}), the optimization problems  (\ref{gmrs12}) and (\ref{gmrs3}) are equivalent so we have the implications
\begin{eqnarray}
\label{gmrs14}
(\ref{gmrs12}) \Leftrightarrow (\ref{gmrs3})  \Rightarrow (\ref{mrs2}).
\end{eqnarray}
Therefore, a maximum of $S[\overline{\omega}]$ at fixed energy and
circulation is a MRS equilibrium state, but the reciprocal is wrong in
case of ensemble inequivalence.  For example, if the MRS equilibrium
vorticity distribution is Gaussian (which corresponds to specific
initial conditions), the generalized entropy $S[\overline{\omega}]$ is
proportional to minus the coarse-grained enstrophy
$\Gamma_2^{c.g.}=\int\overline{\omega}^2\, d{\bf r}$ (see
Sec. \ref{sec_gp}). Therefore, a minimum of coarse-grained enstrophy
at fixed energy and circulation is a MRS equilibrium state, but the
reciprocal is wrong in case of ensemble inequivalence.

{\it Remark 1:} In case of equivalence between microcanonical and grand microcanonical ensembles, these results justify a ``generalized selective decay principle'' \cite{proc}. Indeed, in that case,  the equilibrium  coarse-grained vorticity field maximizes a generalized entropy $S[\overline{\omega}]$ (or minimizes the functional $-S$) at fixed circulation and energy. For a Gaussian equilibrium vorticity distribution, this justifies a minimum coarse-grained enstrophy principle through statistical mechanics. In the present case, the increase of generalized entropy is due to coarse-graining: the microscopic Casimirs $\overline{S[\omega]}$ calculated with the fine-grained vorticity are conserved while the macroscopic Casimirs $S[\overline{\omega}]$ calculated with the coarse-grained vorticity increase. By constrast, the energy $E[\overline{\omega}]$ and the circulation  $\Gamma[\overline{\omega}]$ calculated with the coarse-grained vorticity remain approximately conserved.

{\it Remark 2:} Since  (\ref{mrs2}) is not equivalent to (\ref{gmrs12}), a MRS statistical equilibrium state does not necessarily satisfy the condition of refined dynamical stability (\ref{gmrs12}) given by Ellis-Haven-Turkington \cite{ellis}. However, it can be shown that a MRS statistical equilibrium state is always nonlinearly dynamically stable with respect to the 2D Euler equations as a consequence of the Kelvin-Arnol'd theorem which provides an even more refined condition of nonlinear stability than the EHT criterion (see \cite{proc} for details).

\section{Relaxation equations with a prior vorticity distribution}

Let us now derive some relaxation equations associated with the EHT approach. These relaxation equations will be compared to those associated with the MRS theory.

\subsection{A first type of relaxation equations}
\label{sec_parc}

Chavanis \cite{gen,physicaD,aussois} has proposed a relaxation equation solving the maximization problem (\ref{gmrs3}). This can serve as a numerical algorithm to compute maximum entropy states with appropriate constraints. The idea is to use the two-steps method presented in Sec. \ref{sec_get}. The discussion is here slightly improved.  We assume that, at any time of the evolution, the vorticity distribution $\rho({\bf r},\sigma,t)$ maximizes the relative entropy (\ref{gmrs2}) with the constraint on mean vorticity $\overline{\omega}({\bf r},t)=\int \rho\sigma\, d\sigma$ and normalization $\int \rho\, d\sigma=1$. This leads to the time dependent distribution
\begin{eqnarray}
\rho({\bf r},\sigma,t)=\frac{1}{Z({\bf
r},t)}\chi(\sigma)e^{-\sigma\Phi({\bf r},t)},\label{nge1}
\end{eqnarray}
where $Z({\bf r},t)$ and $\Phi({\bf r},t)$ are determined by
\begin{eqnarray}
Z({\bf r},t)=\int \chi(\sigma)e^{-\sigma\Phi({\bf
r},t)}d\sigma\equiv \hat{\chi}(\Phi),\label{nge2}
\end{eqnarray}
\begin{eqnarray}
\overline{\omega}({\bf r},t)=\frac{1}{Z({\bf r},t)}\int \chi(\sigma)\sigma
e^{-\sigma\Phi({\bf r},t)}d\sigma=-(\ln\hat{\chi})'(\Phi).\nonumber\\\label{nge3}
\end{eqnarray}
Now, according to (\ref{gmrs12}), we know that the equilibrium vorticity field  $\overline{\omega}({\bf r})$ maximizes the generalized entropy
\begin{eqnarray}
\label{genentbx}
S[\overline{\omega}]=-\int C(\overline{\omega}) \, d{\bf r},
\end{eqnarray}
with
\begin{eqnarray}
C(\overline{\omega})=-\int^{\overline{\omega}}
[(\ln\hat\chi)']^{-1}(-x)dx,\label{ge8bx}
\end{eqnarray}
at fixed circulation and energy. We can obtain a relaxation equation for $\overline{\omega}({\bf r},t)$ solving this maximization problem by using a generalized Maximum Entropy Production Principle \cite{gen,physicaD,aussois}. We assume that the coarse-grained vorticity evolves in time so as to maximize the rate of (generalized) entropy production $S[\overline{\omega}]$ (fixed by the prior) at fixed circulation and energy. This leads to a generalized Fokker-Planck equation of the form
\begin{equation}
{\partial \overline{\omega}\over\partial t}+{\bf u}\cdot \nabla
\overline{\omega}=\nabla\cdot \biggl \lbrace D
\biggl\lbrack \nabla\overline{\omega}+{\beta(t)\over
C''(\overline{\omega})}\nabla\psi\biggr\rbrack\biggr\rbrace ,
\label{intro8}
\end{equation}
with
\begin{equation}
\beta(t)=-{\int D\nabla\overline{\omega}\cdot\nabla\psi d{\bf r}\over \int D{(\nabla\psi)^{2}\over C''(\overline{\omega})}d{\bf r}},
\label{intro9}
\end{equation}
where $\beta(t)$ is a Lagrange multiplier enforcing the energy constraint $\dot E=0$ at any time and $D({\bf r},t)$ is the diffusion coefficient. The boundary conditions are
\begin{equation}
\left (\nabla\overline{\omega}+{\beta(t)\over
C''(\overline{\omega})}\nabla\psi\right )\cdot {\bf n}=0,
\label{intro9df}
\end{equation}
where ${\bf n}$ is a unit vector normal to the boundary, in order to guarantee the conservation of circulation. Easy calculations give the generalized $H$-theorem
\begin{equation}
\dot S=\int D C''(\overline{\omega})\left (\nabla\overline{\omega}+\frac{\beta(t)}{C''(\overline{\omega})}\nabla\psi\right )^2\, d{\bf r}\ge 0,
\end{equation}
provided that $D({\bf r},t)\ge 0$. By construction, the relaxed vorticity field $\overline{\omega}({\bf r})$ solves the
maximization problem (\ref{gmrs12}). Then, the corresponding
distribution (\ref{nge1})-(\ref{nge3}) solves the maximization problem
(\ref{gmrs3}). Therefore, these relaxation equations tend to the statistical equilibrium state corresponding to the EHT approach. The diffusion coefficient is not given by the MEPP, but it can be estimated by Eq. (\ref{mrs11d}). Using Eqs. (\ref{nge1})-(\ref{nge3}) and repeating the steps (\ref{idp})-(\ref{ind}), we establish that at any time\footnote{The present theory predicts that the local centered variance $\omega_2({\bf r},t)$ is correlated to the coarse-grained vorticity $\overline{\omega}({\bf r},t)$ by Eq. (\ref{corr}) even in the out-of-equilibrium regime. It could be interesting to confront this prediction with experiments or  observations in the oceans and in the atmosphere.}:
\begin{eqnarray}
\omega_2=\frac{1}{C''(\overline{\omega})}.
\label{corr}
\end{eqnarray}
Plugging this result in Eq. (\ref{mrs11d}), we obtain the expression of the diffusion coefficient
\begin{equation}
D({\bf r},t)=\frac{K\epsilon^2}{\sqrt{C''(\overline{\omega})}}.\label{di}
\end{equation}
Finally, using arguments similar to those of Sec. \ref{sec_nept}, the equation for the velocity field is
\begin{eqnarray}
\label{nept5b}
\frac{\partial {\bf u}}{\partial t}+({\bf u}\cdot \nabla) {\bf u}=-\frac{1}{\rho}\nabla p+   D\left (\Delta {\bf u}-\frac{\beta(t)}{C''(\overline{\omega})}{\bf u}\right ).
\end{eqnarray}
At equilibrium, we get
\begin{eqnarray}
\label{nept6b}
\Delta {\bf u}=\frac{\beta}{C''(\overline{\omega})}{\bf u},
\end{eqnarray}
which can be directly derived from the Gibbs state (\ref{mrs4}) using $\Delta {\bf u}={\bf z}\times \nabla\overline{\omega}$, $\nabla\overline{\omega}=\overline{\omega}'(\psi)\nabla\psi$ and Eq. (\ref{css}).

Therefore, the system of equations (\ref{nge1})-(\ref{di}) is
completely closed when the prior vorticity distribution $\chi(\sigma)$
is given\footnote{The relaxation equation (\ref{intro8}) can also be
obtained from the first moment (\ref{mrs11}) of the relaxation
equations (\ref{mrs8})-(\ref{mrs9}) by using the identity
(\ref{corr}).}. Note that they determine not only the evolution of
the mean flow $\overline{\omega}({\bf r},t)$ through
Eqs. (\ref{intro8}), (\ref{intro9}) and (\ref{di}) but also the
evolution of the full vorticity distribution $\rho({\bf r},\sigma,t)$
through Eqs. (\ref{nge1}), (\ref{nge2}) and (\ref{nge3}).  The case of
a Gaussian prior will be discussed specifically in
Sec. \ref{sec_gp}. However, we stress that the relaxation equations (\ref{nge1})-(\ref{di}), which are relatively easy to solve numerically, can be
used to study situations going beyond the Gaussian approximation. Some
examples showing the construction of the generalized entropy
$S[\overline{\omega}]$ from the prior $\chi(\sigma)$ and giving the
corresponding equilibrium states and the corresponding relaxation
equations are discussed in
\cite{gen,physicaD,aussois}. Apart from their
potential interest in 2D turbulence, this leads to interesting
classes of nonlinear mean field Fokker-Planck equations \cite{nfp}.

It should be emphasized that the relaxation equations associated with the maximization problem (\ref{gmrs12}) are not unique. For example, another type of relaxation equations (see Appendix \ref{sec_der}) solving the maximization problem (\ref{gmrs12}) is given by \cite{proc}:
\begin{equation}
\label{two22new}
\frac{\partial\overline{\omega}}{\partial t}+{\bf u}\cdot \nabla\overline{\omega}=-D(C'(\overline{\omega})+\beta(t)\psi+\alpha(t)),
\end{equation}
where $D$ is a positive coefficient and the Lagrange multipliers $\beta(t)$ and $\alpha(t)$ evolve  according to
\begin{equation}
\label{two23new}
\langle DC'(\overline{\omega})\psi\rangle+\beta(t)\langle D\psi^{2}\rangle+\alpha(t)\langle D\psi\rangle=0,
\end{equation}
\begin{equation}
\label{two24new}
\langle DC'(\overline{\omega})\rangle+\beta(t)\langle D\psi\rangle+\alpha(t)\langle D\rangle=0,
\end{equation}
so as to conserve the energy and the circulation (the brackets represent the domain average $\langle X\rangle=\int X\, d{\bf r}$). The boundary conditions are
\begin{equation}
\label{two24newgod}
C'(\overline{\omega})=-\alpha(t),
\end{equation}
on the boundary, in order to be consistent with the equilibrium state
where the r.h.s. of Eq. (\ref{two22new}) is equal to
zero on the whole domain (recall that $\psi=0$ on the boundary). Easy
calculations lead to the generalized $H$-theorem
\begin{equation}
\dot S=\int D(C'(\overline{\omega})+\beta(t)\psi+\alpha(t))^2\, d{\bf r}\ge 0.\label{hnm}
\end{equation}
The relaxed vorticity field $\overline{\omega}({\bf r})$ solves the
maximization problem (\ref{gmrs12}). Then, the corresponding
distribution (\ref{nge1})-(\ref{nge3}) solves the maximization problem
(\ref{gmrs3}).

{\it Remark:} Since the maximization problem (\ref{gmrs12}) provides a refined criterion of nonlinear dynamical stability for the 2D Euler equations \cite{ellis}, the relaxation equations of this section can also be used as numerical algorithms to construct nonlinearly dynamically stable steady states of the 2D Euler equation independently from the statistical mechanics interpretation \cite{proc}.

\subsection{A new type of relaxation equations}
\label{sec_pn}

In the previous approach, the vorticity distribution is assumed to have the form  (\ref{nge1})-(\ref{nge3}) at each time. This assumption is consistent with the two-steps method developed in Sec. \ref{sec_get} to show the equivalence between the maximization problems (\ref{gmrs3}) and (\ref{gmrs12}). We shall now introduce a new type of relaxation equations in which the form of the vorticity distribution changes with time \cite{proc}. These relaxation equations are associated with the basic maximization problem (\ref{gmrs3}).

We write the  equation for the evolution of the vorticity distribution $\rho({\bf r},\sigma,t)$ in the  form
\begin{equation}
\label{na1}
\frac{\partial\rho}{\partial t}+{\bf u}\cdot\nabla\rho=-\frac{\partial J}{\partial \sigma},
\end{equation}
where $J({\bf r},\sigma,t)$ is a current acting in the space of vorticity levels $\sigma$. It will be determined by a systematic procedure. We note that this form assures the conservation of the local normalization $\int\rho d\sigma=1$ provided that $J\rightarrow 0$ for $\sigma\rightarrow \pm\infty$. We also emphasize that the total areas of the vorticity levels $\gamma(\sigma,t)=\int \rho d{\bf r}$ are {\it not} conserved by the relaxation equations (\ref{na1}). This is because, in the EHT approach, the Casimirs are not conserved. This differs from the relaxation equations (\ref{mrs8}) associated with the MRS approach where the left hand side is of the form $-\nabla\cdot {\bf J}$ where ${\bf J}({\bf r},\sigma,t)$ is a current acting in position space. In the MRS approach, the Casimirs, or equivalently the total areas of the vorticity levels $\gamma(\sigma,t)=\int \rho d{\bf r}$, are conserved by the relaxation equations. This is not the case in the EHT approach. Therefore, we see from the start that the structure of the relaxation equations will be very different in the two approaches.

Multiplying Eq. (\ref{na1}) by $\sigma$ and integrating on the vorticity levels, we obtain an equation for the coarse-grained vorticity
\begin{equation}
\label{na2}
\frac{\partial\overline{\omega}}{\partial t}+{\bf u}\cdot\nabla\overline{\omega}= \int Jd\sigma,
\end{equation}
where we have used an integration by parts to get the r.h.s. We can also derive an equation giving the evolution of the local centered variance $\omega_2$. Multiplying Eq. (\ref{na1}) by $\sigma^2$ and integrating on $\sigma$, we obtain
\begin{equation}
\label{na5}
\frac{\partial\overline{\omega^2}}{\partial t}+{\bf u}\cdot\nabla\overline{\omega^2}=2 \int J\sigma\, d\sigma,
\end{equation}
where we  have used an integration by parts to get the r.h.s. On the other hand, multiplying Eq. (\ref{na2})  by $\overline{\omega}$, we get
\begin{equation}
\label{na6}
\frac{\partial\overline{\omega}^2}{\partial t}+{\bf u}\cdot\nabla\overline{\omega}^2=2\overline{\omega} \int J d\sigma.
\end{equation}
Subtracting these two equations, we find that
\begin{equation}
\label{na7}
\frac{\partial {\omega}_2}{\partial t}+{\bf u}\cdot\nabla {\omega}_2=2 \int J(\sigma-\overline{\omega}) d\sigma.
\end{equation}

Let us now consider the constraints that the relaxation equations must satisfy. The conservation of the circulation implies
\begin{equation}
\label{na3}
\dot\Gamma=\int J d{\bf r}d\sigma=0,
\end{equation}
and the conservation of the energy implies
\begin{equation}
\label{na4}
\dot E=\int J\psi d{\bf r}d\sigma=0.
\end{equation}
On the other hand, the rate of production of relative entropy $S_\chi$ is given by
\begin{equation}
\label{na8}
\dot S_{\chi}=-\int J \frac{\partial}{\partial\sigma}\ln\left \lbrack\frac{\rho}{\chi(\sigma)}\right\rbrack d{\bf r}d\sigma.
\end{equation}
We shall determine the optimal current $J$ by maximizing the rate of relative entropy production at fixed circulation and energy. We also introduce the physical constraint $\int J^2/(2\rho)d\sigma\le C$ that puts a bound on the current. We write the variational problem as
\begin{equation}
\label{na9}
\delta\dot S_{\chi}-\beta(t)\delta\dot E-\alpha(t)\delta\dot\Gamma-\int \frac{1}{D}\delta\left (\frac{J^{2}}{2\rho}d\sigma\right )\, d{\bf r}=0,
\end{equation}
where $\alpha(t)$, $\beta(t)$ and $D({\bf r},t)$  are Lagrange multipliers that assure the conservation of circulation and energy at any time. Performing the variations, we find that the optimal current is
\begin{equation}
\label{na10}
J=-D\rho\left\lbrace \frac{\partial}{\partial\sigma}\ln\left\lbrack \frac{\rho}{\chi(\sigma)}\right\rbrack+\beta(t)\psi+\alpha(t)\right\rbrace,
\end{equation}
or equivalently
\begin{equation}
\label{na11}
J=-D\left\lbrace \frac{\partial\rho}{\partial\sigma}-\rho (\ln\chi)'(\sigma)+\beta(t)\rho\psi+\alpha(t)\rho\right\rbrace.
\end{equation}
Therefore, the relaxation equation for the vorticity distribution $\rho({\bf r},\sigma,t)$ takes the form
\begin{eqnarray}
\label{na12}
\frac{\partial\rho}{\partial t}+{\bf u}\cdot\nabla\rho\qquad\qquad\qquad\qquad\qquad\qquad\qquad\qquad\nonumber\\
=\frac{\partial}{\partial\sigma}\left\lbrack D\left\lbrace \frac{\partial\rho}{\partial\sigma}-\rho (\ln\chi)'(\sigma)+\beta(t)\rho\psi+\alpha(t)\rho\right\rbrace\right\rbrack.
\end{eqnarray}
Using Eq. (\ref{na2}), we obtain the relaxation equation for the coarse-grained vorticity
\begin{eqnarray}
\label{na13}
\frac{\partial\overline{\omega}}{\partial t}+{\bf u}\cdot\nabla\overline{\omega}=D\left \lbrack \int \rho (\ln\chi)'(\sigma)d\sigma-\beta(t)\psi-\alpha(t)\right \rbrack. \nonumber\\
\end{eqnarray}
In general this equation is not closed as it depends on the full vorticity distribution $\rho({\bf r},\sigma,t)$. The Lagrange multipliers are determined by Eqs. (\ref{na3}), (\ref{na4}) and (\ref{na11}) yielding
\begin{eqnarray}
\label{na14}
\int \langle D\rho\rangle (\ln\chi)'(\sigma)d\sigma-\beta(t)\langle D\psi\rangle-\alpha(t)\langle D\rangle=0,
\end{eqnarray}
\begin{eqnarray}
\label{na15}
\int \langle D\rho\psi\rangle(\ln\chi)'(\sigma)d\sigma-\beta(t)\langle D\psi^2\rangle-\alpha(t)\langle D\psi\rangle=0.\quad
\end{eqnarray}
Substituting Eq. (\ref{na10}) in Eq. (\ref{na8}) and using the constraints (\ref{na3})-(\ref{na4}), it is easy to establish the H-theorem
\begin{eqnarray}
\label{na16}
\dot S_{\chi}=\int \frac{J^{2}}{D\rho}d{\bf r}d\sigma\ge 0,
\end{eqnarray}
provided that $D\ge 0$. On the other hand, a stationary solution corresponds to $J=0$ yielding
\begin{eqnarray}
\label{na17}
\frac{\partial}{\partial\sigma}\ln\left\lbrack \frac{\rho({\bf r},\sigma)}{\chi(\sigma)}\right\rbrack+\beta\psi+\alpha=0.
\end{eqnarray}
After integration, we recover the Gibbs state (\ref{mrs4}).
Equation (\ref{na12}) with the constraints
(\ref{na14})-(\ref{na15}) has the following properties: (i)
$\Gamma$, $E$ and $\int
\rho d\sigma=1$ are conserved. (ii) $\dot S_{\chi}\ge 0$. (iii) $\dot
S_{\chi}=0$ $\Leftrightarrow$ $\rho({\bf r},\sigma)$ is the Gibbs
state (\ref{mrs4}) $\Leftrightarrow$ $\partial_{t}\rho=0$.  (iv)
$\rho({\bf r},\sigma)$ is a steady state of
Eqs. (\ref{na12}),
(\ref{na14}), (\ref{na15}) iff it is a critical point of
$S_{\chi}$ at fixed $\Gamma$, $E$ and normalization. (v) a steady
state of Eqs. (\ref{na12}),
(\ref{na14}), (\ref{na15}) is linearly stable iff it
is a (local) maximum of $S_{\chi}$ at fixed $\Gamma$, $E$ and
normalization. By Lyapunov's direct method, we know that if $S_{\chi}$
is bounded from above, Eqs.  (\ref{na12}),
(\ref{na14}), (\ref{na15})  will relax
towards a (local) maximum of $S_{\chi}$ at fixed $\Gamma$, $E$ and
normalization (if several local maxima exist, the choice of the
maximum will depend on a notion of basin of attraction). Therefore, a
stable steady state of Eqs. (\ref{na12}),
(\ref{na14}), (\ref{na15})   solves the
maximization problem (\ref{gmrs3}) for a given prior and a given
circulation and energy specified by the initial condition.

We can easily derive a hierarchy of equations for the local moments of the vorticity.
Multiplying Eq. (\ref{na1}) by $\sigma^n$ and integrating on $\sigma$, we get
\begin{equation}
\label{hier1}
\frac{\partial\overline{\omega^n}}{\partial t}+{\bf u}\cdot\nabla\overline{\omega^n}=n \int J\sigma^{n-1}\, d\sigma.
\end{equation}
Inserting the expression (\ref{na11}) of the current, we obtain
\begin{eqnarray}
\label{hier2}
\frac{\partial\overline{\omega^n}}{\partial t}+{\bf u}\cdot\nabla\overline{\omega^n}=Dn(n-1)\overline{\omega^{n-2}}\nonumber\\
+Dn\biggl\lbrace \int\rho (\ln\chi)'(\sigma)\sigma^{n-1}\, d\sigma\nonumber\\
-\beta(t)\overline{\omega^{n-1}}\psi-\alpha(t)\overline{\omega^{n-1}}\biggr\rbrace.
\end{eqnarray}
If we recall that
\begin{eqnarray}
\label{hier3}
\ln\chi(\sigma)=-\sum_{k>1}\alpha_k\sigma^k,
\end{eqnarray}
we can rewrite the foregoing equation in the form
\begin{eqnarray}
\label{hier4}
\frac{\partial\overline{\omega^n}}{\partial t}+{\bf u}\cdot\nabla\overline{\omega^n}=Dn(n-1)\overline{\omega^{n-2}}\nonumber\\
-Dn\biggl\lbrace \sum_{k>1}\alpha_k k \overline{\omega^{k+n-2}}
+\beta(t)\overline{\omega^{n-1}}\psi+\alpha(t)\overline{\omega^{n-1}}\biggr\rbrace.\quad
\end{eqnarray}
From this general expression, we see that the hierarchy of equations is closed only if $a_k=0$ for $k>2$ that is to say for a Gaussian prior (see next section). Otherwise, the equation for the moment of order $n$ requires the knowledge of the moment of order $n+(k-2)>n$ and some closure approximations must be introduced. This can be an interesting mathematical problem but it will not be considered in this paper.

{\it Remark:} According to Eq. (\ref{gmrs9}), the relaxation equations presented in this section also solve the dual maximization problem (\ref{mrs2}) for the corresponding values of the Casimirs. Therefore, according to the interpretation of Sec. \ref{sec_other} they can be used as numerical algorithms to construct a subclass of MRS statistical
equilibria.

\section{The case of a Gaussian prior}
\label{sec_gp}

In this section, we consider the particular  case of a Gaussian prior vorticity distribution and make the connection with minimum enstrophy states.

\subsection{Equilibrium states}

We assume a Gaussian prior of the form
\begin{eqnarray}
\label{g1}
\chi(\sigma)=\frac{1}{\sqrt{2\pi\Omega_2}}e^{-\frac{\sigma^2}{2\Omega_2}},
\end{eqnarray}
where $\Omega_2$ is a constant. The corresponding Gibbs state (\ref{mrs4}) is
\begin{equation}
\label{gg2}
\rho({\bf r},\sigma)=\frac{1}{Z}\frac{1}{\sqrt{2\pi\Omega_2}}
e^{-\frac{\sigma^2}{2\Omega_2}}e^{-\sigma(\beta\psi+\alpha)}.
\end{equation}
The vorticity and the local centered variance are given by
\begin{equation}
\label{op}
\overline{\omega}({\bf r})=-\Omega_2(\beta\psi({\bf r})+\alpha),
\end{equation}
and
\begin{equation}
{\omega}_2({\bf r})=\Omega_2.
\end{equation}
Therefore, in the case of a Gaussian prior, the $\overline{\omega}-\psi$ relationship is linear and the local centered variance is uniform. The Gibbs state (\ref{gg2}) can be rewritten
\begin{equation}
\label{re}
\rho({\bf r},\sigma)=\frac{1}{\sqrt{2\pi\Omega_2}}e^{-\frac{(\sigma-\overline{\omega})^2}{2\Omega_2}}.
\end{equation}
At equilibrium, substituting Eq. (\ref{re}) in Eqs. (\ref{mrs1}) and (\ref{gmrs2}), we get $S[\rho_{eq}]=\frac{1}{2}\ln(\Gamma_2^{f.g.}-\Gamma_2^{c.g.})$ and $S_\chi[\rho_{eq}]=-\frac{1}{2\Omega_2}\Gamma_2^{c.g.}$ (up to additive constants). On the other hand, the generalized entropy defined by Eqs.  (\ref{ge6}) and (\ref{ge8}) can be conveniently  calculated from Eqs. (\ref{cba}) and (\ref{op}) yielding $C'(\overline{\omega})=\overline{\omega}/\Omega_2$, hence
\begin{equation}
\label{g10}
S=-\int \frac{\overline{\omega}^{2}}{2\Omega_2}d{\bf r}=-\frac{1}{2\Omega_2}\Gamma_2^{c.g.}.
\end{equation}
Therefore, for a Gaussian prior, the generalized entropy is proportional to minus the macroscopic enstrophy $\Gamma_2^{c.g.}=\int \overline{\omega}^{2}\, d{\bf r}$. In that case, according to Eq. (\ref{eqa}), the maximization of the relative entropy $S_{\chi}$ at fixed energy and circulation (EHT thermodynamical stability) is equivalent to the minimization of the macroscopic enstrophy $\Gamma_2^{c.g.}$ at fixed energy and circulation, i.e.
\begin{eqnarray}
\label{minen}
\min_{\overline{\omega}}\quad \lbrace \Gamma_2^{c.g.}[\overline{\omega}]\quad |\quad \Gamma[\overline{\omega}]=\Gamma, \quad E[\overline{\omega}]=E \rbrace.
\end{eqnarray}
According to the interpretation given in Sec. \ref{sec_other}, this minimization problem also provides a {\it sufficient} condition of MRS thermodynamical stability.

{\it Remark}: Writing the variational problem in the form (\ref{cbb}), the critical points of $S$, given by Eq. (\ref{g10}), at fixed energy and circulation are given by Eq. (\ref{op}). Furthermore, they are maxima of $S$ at fixed energy and circulation iff
\begin{eqnarray}
\label{jkg}
\delta^2J=-\frac{1}{2\Omega_2}\int (\delta\overline{\omega})^2\, d{\bf r}-\frac{1}{2}\beta\int\delta\overline{\omega}\delta\psi\, d{\bf r}<0,
\end{eqnarray}
for all perturbations $\delta\overline{\omega}$ that conserve energy
and circulation at first order.

\subsection{A first type of relaxation equations}

We first consider the relaxation equations of Sec. \ref{sec_parc}. In this approach, for a Gaussian prior, the vorticity distribution is at any time given by
\begin{equation}
\rho({\bf r},\sigma,t)=\frac{1}{\sqrt{2\pi\Omega_2}}e^{-\frac{(\sigma-\overline{\omega}({\bf r},t))^2}{2\Omega_2}}.
\end{equation}
Therefore, the vorticity distribution is Gaussian and the local centered variance $\omega_2({\bf r},t)=\Omega_2$ is uniform and constant in time. Using $C''(\overline{\omega})=1/\Omega_2$ according to Eq. (\ref{g10}), the evolution of the vorticity is given by
\begin{equation}
{\partial \overline{\omega}\over\partial t}+{\bf u}\cdot \nabla
\overline{\omega}=\nabla\cdot \biggl \lbrace D
\biggl\lbrack \nabla\overline{\omega}+\beta(t)\Omega_2\nabla\psi\biggr\rbrack\biggr\rbrace ,
\label{intro8n}
\end{equation}
\begin{equation}
\beta(t)=-{\int D\nabla\overline{\omega}\cdot\nabla\psi d{\bf r}\over \int D \Omega_2 (\nabla\psi)^{2}d{\bf r}},
\label{intro9n}
\end{equation}
\begin{equation}
D={K\epsilon^2}\Omega_2.
\end{equation}
Since the  diffusion coefficient $D$ is constant, Eqs. (\ref{intro8n}) and (\ref{intro9n}) can also be written
\begin{equation}
{\partial \overline{\omega}\over\partial t}+{\bf u}\cdot \nabla
\overline{\omega}=D(\Delta\overline{\omega}-\beta(t)\Omega_2\overline{\omega}),
\label{intro8b}
\end{equation}
\begin{equation}
\beta(t)={-\int \overline{\omega}^2\, d{\bf r}\over \Omega_2\int (\nabla\psi)^{2}d{\bf r}}=\frac{-\Gamma_2^{c.g.}(t)}{2\Omega_2 E}=\frac{S(t)}{E},
\label{intro9b}
\end{equation}
where we have used an integration by parts to simplify the last equation. Interestingly, for this Gaussian model, we see that $S(t)=\beta(t)E$ out-of-equilibrium. On the other hand, the $H$ theorem can be written
\begin{eqnarray}
\dot S=-\frac{1}{2\Omega_2}\dot{\Gamma}_2^{c.g.}=\int \frac{D}{\Omega_2} \left (\nabla\overline{\omega}+\beta(t)\Omega_2\nabla\psi\right )^2\, d{\bf r}\ge 0.\nonumber\\
\end{eqnarray}
Therefore, the relaxation equations increase the generalized entropy (or decrease the coarse-grained enstrophy) at fixed energy and circulation until the maximum entropy (or minimum enstrophy) state is reached. Note also that the equation for the velocity field (\ref{nept5b}) takes the form
\begin{eqnarray}
\label{nept5bb}
\frac{\partial {\bf u}}{\partial t}+({\bf u}\cdot \nabla) {\bf u}=-\frac{1}{\rho}\nabla p+   D\left (\Delta {\bf u}-{\beta(t)}\Omega_2{\bf u}\right ).
\end{eqnarray}
At equilibrium, we get
\begin{eqnarray}
\label{nept6bb}
\Delta {\bf u}={\beta}\Omega_2{\bf u}.
\end{eqnarray}

On the other hand, using $C'(\overline{\omega})=\overline{\omega}/\Omega_2$ according to Eq. (\ref{g10}), the alternative relaxation equations (\ref{two22new})-(\ref{two24new}) become (we here assume $D$ constant to slightly simplify the expressions)
\begin{equation}
\label{two22newb}
\frac{\partial\overline{\omega}}{\partial t}+{\bf u}\cdot \nabla\overline{\omega}=-D\left \lbrack \frac{1}{\Omega_2}\overline{\omega}+\beta(t)\psi+\alpha(t)\right \rbrack,
\end{equation}
\begin{eqnarray}
\label{g7}
\beta(t)=\frac{1}{\Omega_2}\frac{\Gamma\langle\psi\rangle-2AE}{A\langle\psi^2\rangle-\langle\psi\rangle^2},
\end{eqnarray}
\begin{eqnarray}
\label{g8}
\alpha(t)=-\frac{1}{\Omega_2}\frac{\Gamma\langle\psi^2\rangle-2E\langle\psi\rangle}
{A\langle\psi^2\rangle-\langle\psi\rangle^2},
\end{eqnarray}
where $A$ is the domain area.
The $H$-theorem (\ref{hnm}) can be written
\begin{eqnarray}
\dot S=-\frac{1}{2\Omega_2}\dot{\Gamma}_2^{c.g.}=\int D\left (\frac{\overline{\omega}}{\Omega_2}+\beta(t)\psi+\alpha(t)\right )^2\, d{\bf r}\ge 0.\nonumber\\\label{fr}
\end{eqnarray}

\subsection{A new type of relaxation equations} \label{subs:relax_eq}

Let us now consider the new relaxation equations of Sec. \ref{sec_pn}.  Since $(\ln\chi)'(\sigma)=-\sigma/\Omega_2$ for the Gaussian prior (\ref{g1}),
the current (\ref{na11}) becomes
\begin{eqnarray}
\label{g2}
J=- D\left\lbrace \frac{\partial\rho}{\partial\sigma}+\frac{1}{\Omega_2}\rho\sigma
+\beta(t)\rho\psi+\alpha(t)\rho\right\rbrace,
\end{eqnarray}
and the evolution of the vorticity distribution is given by a relaxation equation of the form
\begin{eqnarray}
\label{g3}
\frac{\partial\rho}{\partial t}+{\bf u}\cdot\nabla\rho\qquad\qquad\qquad\qquad\qquad\qquad\qquad\qquad\nonumber\\
=\frac{\partial}{\partial\sigma}\left\lbrack D\left\lbrace \frac{\partial\rho}{\partial\sigma}+\frac{1}{\Omega_2}\rho\sigma
+\beta(t)\rho\psi+\alpha(t)\rho\right\rbrace\right\rbrack.\qquad
\end{eqnarray}
We note that the form of the distribution $\rho({\bf r},\sigma,t)$ changes with time. The vorticity distribution is Gaussian only at equilibrium. The relaxation equation for the coarse-grained vorticity (\ref{na13}) becomes
\begin{equation}
\label{g4}
\frac{\partial\overline{\omega}}{\partial t}+{\bf u}\cdot\nabla\overline{\omega}=-D\left \lbrack \frac{1}{\Omega_2}\overline{\omega}+\beta(t)\psi+\alpha(t)\right \rbrack,
\end{equation}
and the equations (\ref{na14})-(\ref{na15}) for the Lagrange multipliers reduce to (we here assume $D$ constant to slightly simplify the expressions)
\begin{eqnarray}
\label{g7b}
\beta(t)=\frac{1}{\Omega_2}\frac{\Gamma\langle\psi\rangle-2AE}{A\langle\psi^2\rangle-\langle\psi\rangle^2},
\end{eqnarray}
\begin{eqnarray}
\label{g8b}
\alpha(t)=-\frac{1}{\Omega_2}\frac{\Gamma\langle\psi^2\rangle-2E\langle\psi\rangle}{A\langle\psi^2\rangle-\langle\psi\rangle^2}.
\end{eqnarray}
We emphasize that, for a Gaussian prior, the equation for the coarse-grained vorticity (\ref{g4}) is {\it closed}.
Furthermore, it coincides with Eq. (\ref{two22newb}).  Using Eq. (\ref{na7}) and the expression  (\ref{g2}) of the current,  we find that the equation for the local centered variance is
\begin{equation}
\label{g9}
\frac{\partial {\omega}_2}{\partial t}+{\bf u}\cdot\nabla {\omega}_2=2D\left (1-\frac{\omega_2}{\Omega_2}\right ).
\end{equation}
This is also a closed equation. Using Eq. (\ref{hier4}), the relaxation equations for all the local moments of the vorticity are
\begin{eqnarray}
\frac{\partial\overline{\omega^n}}{\partial t}+{\bf u}\cdot\nabla\overline{\omega^n}=Dn(n-1)\overline{\omega^{n-2}}\nonumber\\
-Dn\biggl\lbrack \frac{1}{\Omega_2}\overline{\omega^n}
+\beta(t)\overline{\omega^{n-1}}\psi+\alpha(t)\overline{\omega^{n-1}}\biggr\rbrack.
\end{eqnarray}
Finally, for a Gaussian prior vorticity distribution, we can explicitly check that the generalized entropy (\ref{g10}) increases monotonically for a dynamics of the form (\ref{g3}). Indeed, since Eqs. (\ref{g4}) and (\ref{two22newb}) coincide, we directly obtain the $H$-theorem (\ref{fr}). We have not been able to prove that the generalized entropies $S[\overline{\omega}]$ increase monotonically for a dynamics of the form (\ref{na12}) in the case of a prior that in not Gaussian. This could be an interesting mathematical problem to consider.

{\it Remark:} Note that $D$ plays the role of a diffusion coefficient in the space of vorticity levels [see Eq. (\ref{g3})]. However, in the vorticity equation (\ref{g4}), the quantity $D/\Omega_2\sim 1/t_{relax}$ plays the role of an inverse relaxation time.

\section{Numerical simulations}
\label{sec_num}

\begin{figure}[t]
    \centering \includegraphics[width=4.2cm]{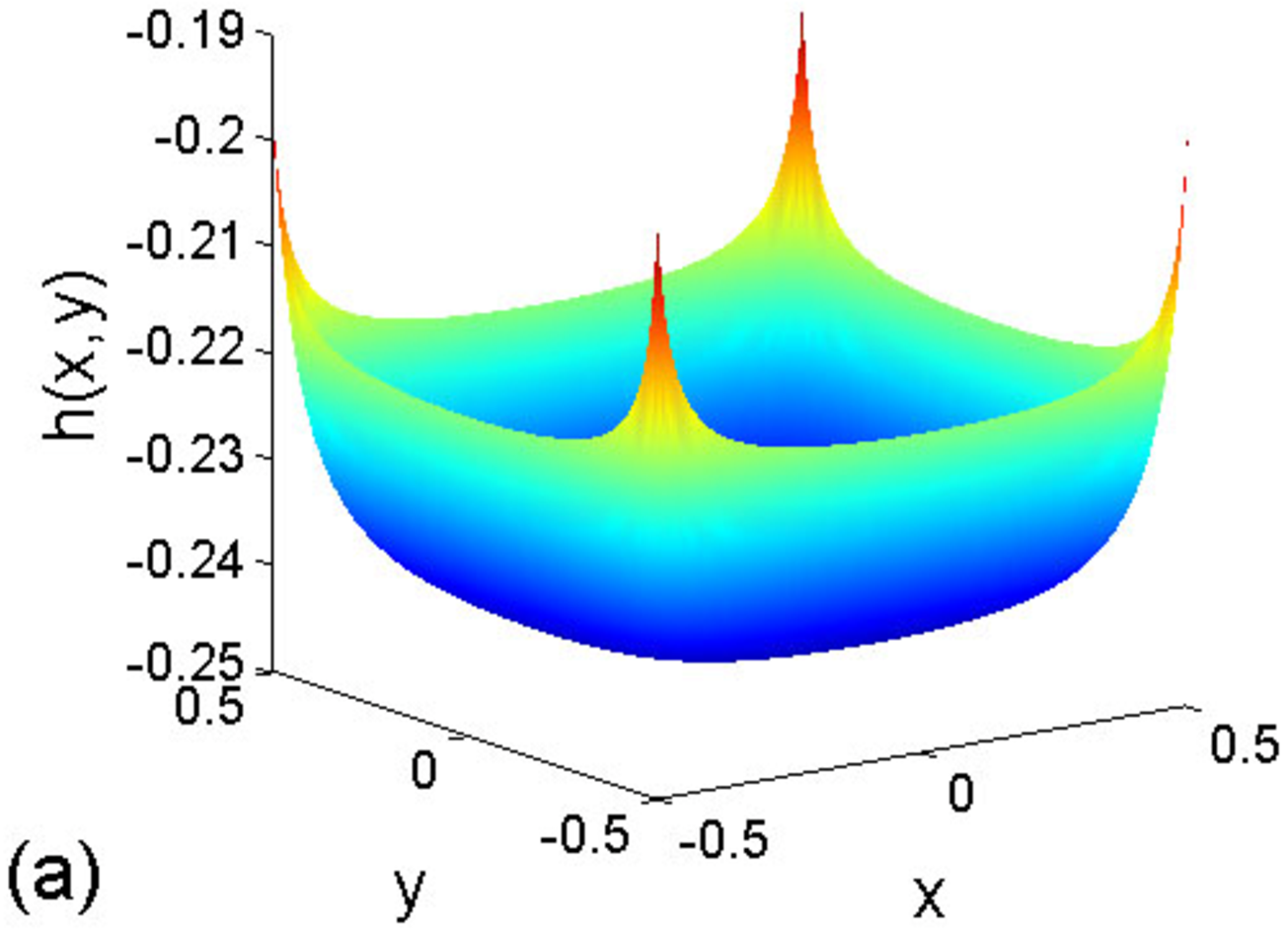}
    \includegraphics[width=4.2cm]{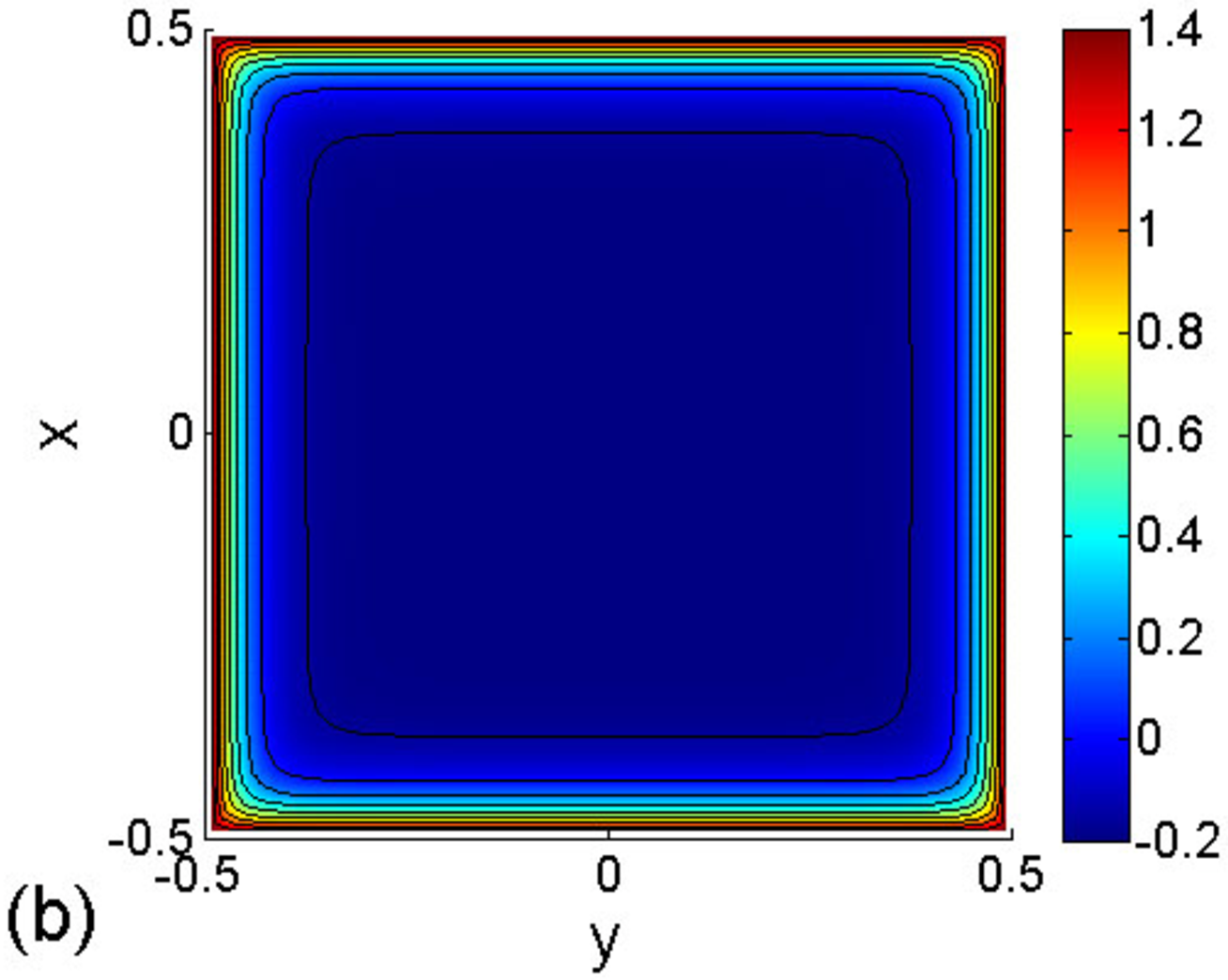}
    \includegraphics[width=4.2cm]{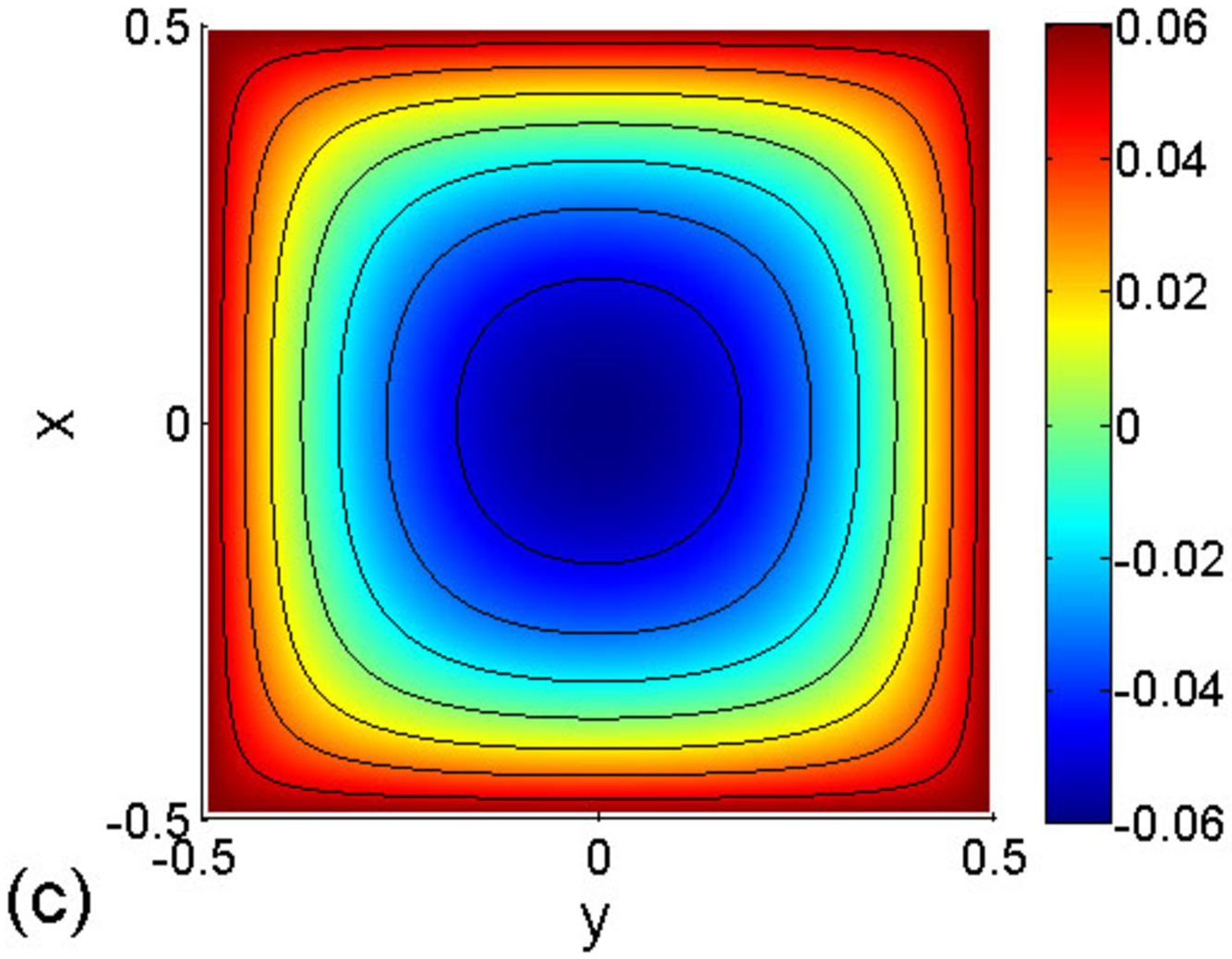}
    \includegraphics[width=4.2cm]{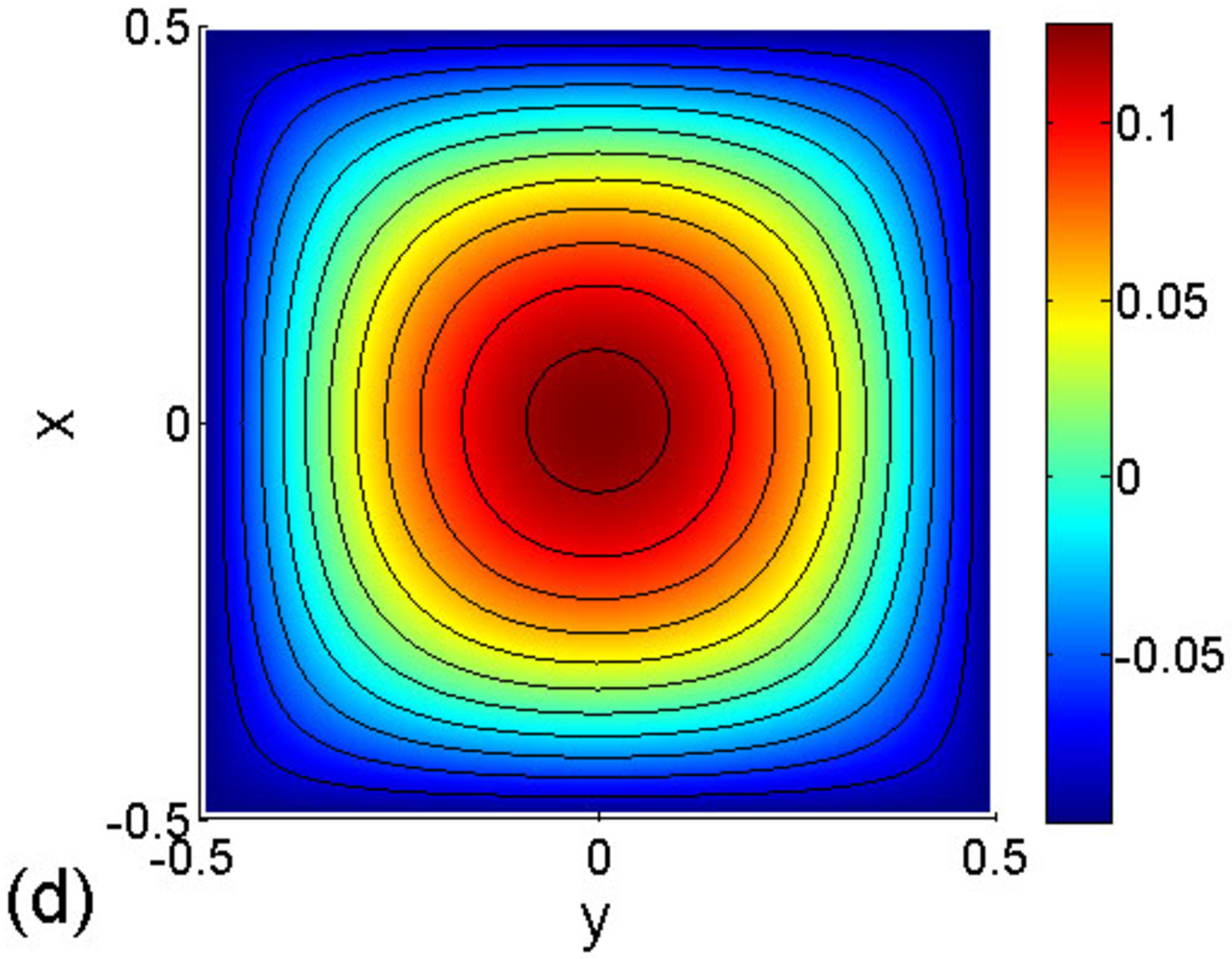} \caption{Continental
    slope: (a) topography $h(x,y)$; (b) to (d)
    Equilibrium PV field for (b) $\beta=1000$, (c)
    $\beta=10$, (d) $\beta=-10$.}  \label{fig:cont_slope_b}
\end{figure}

\begin{figure}[b]
    \centering
        \includegraphics[width=4.2cm]{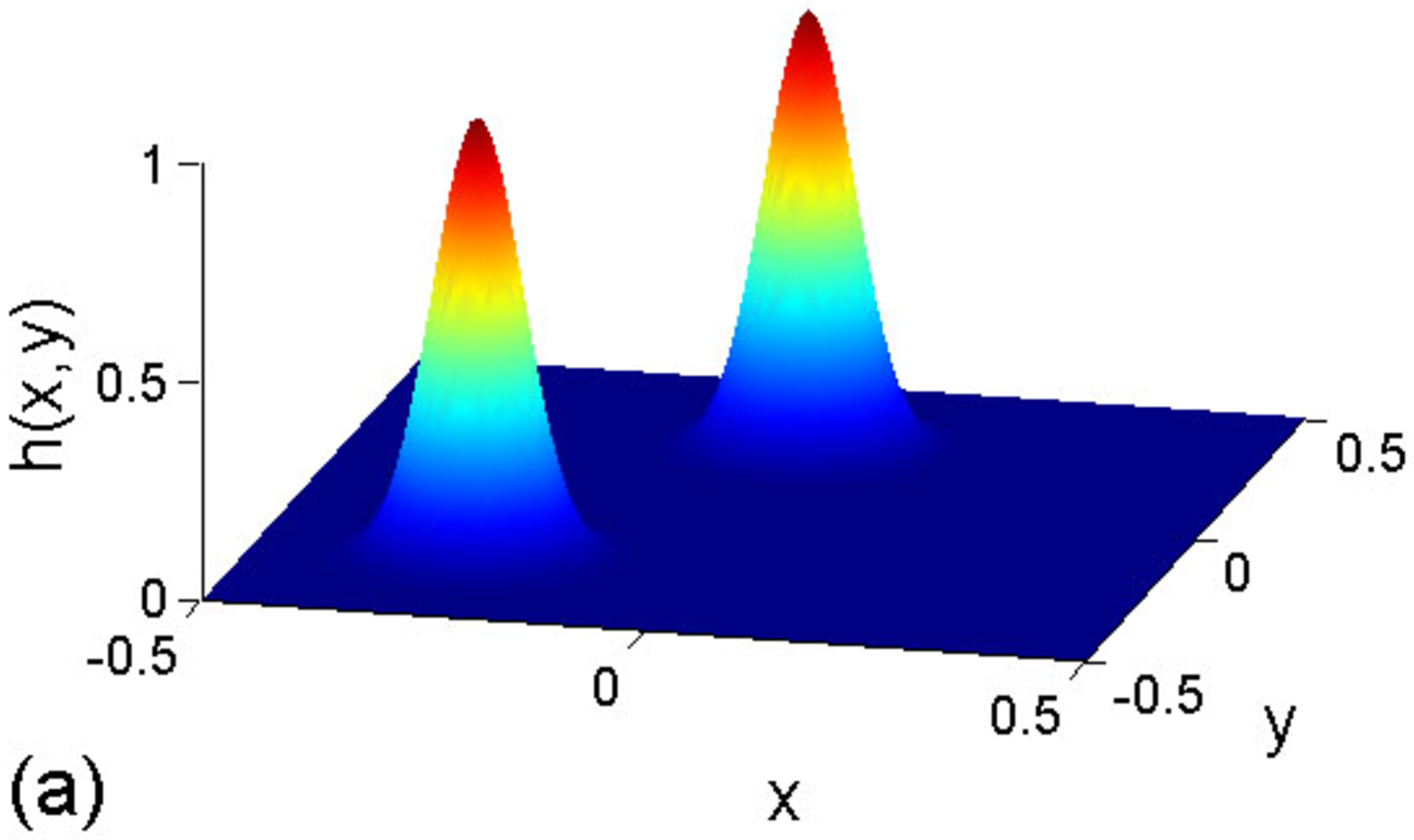}
        \includegraphics[width=4.2cm]{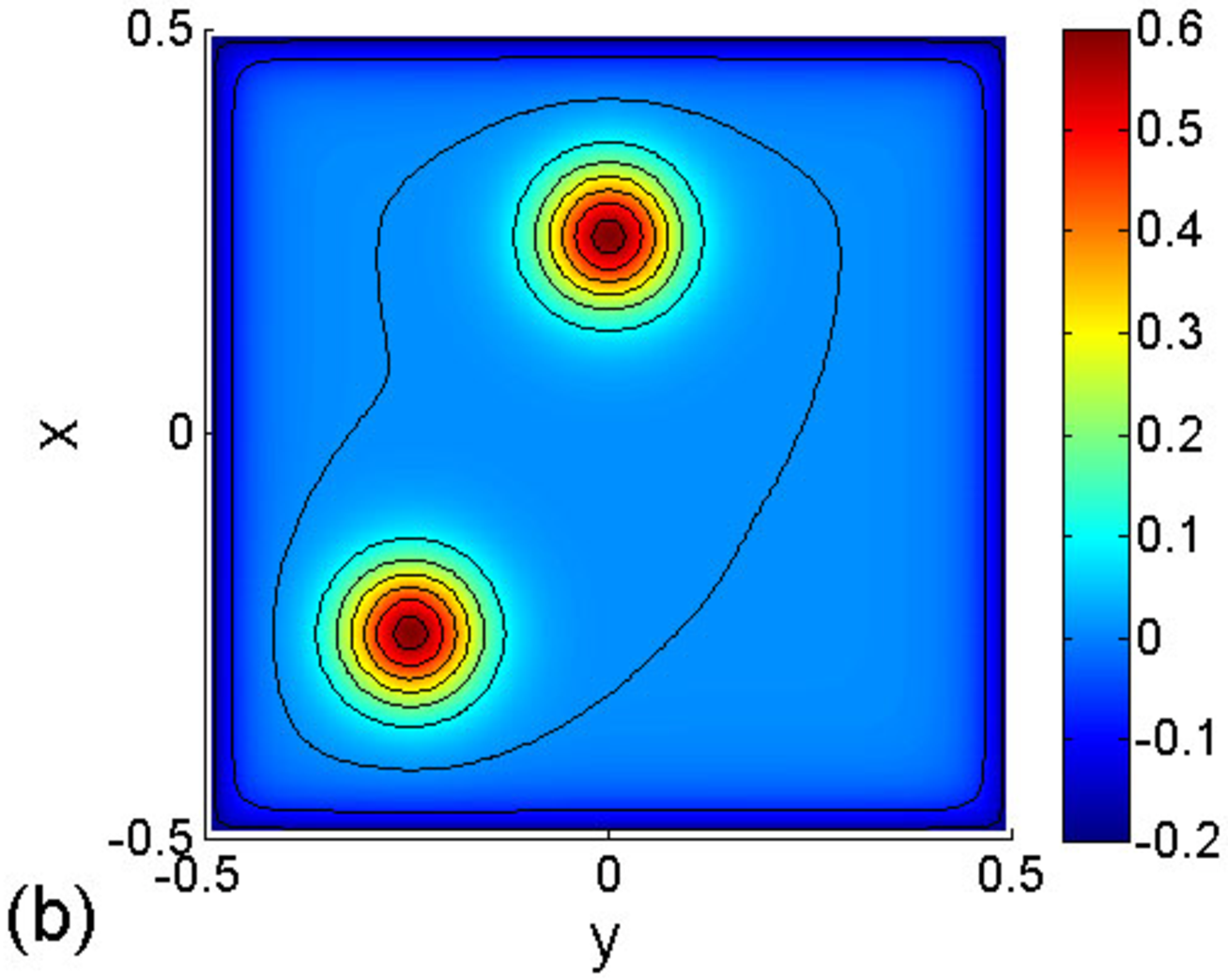}
        \includegraphics[width=4.2cm]{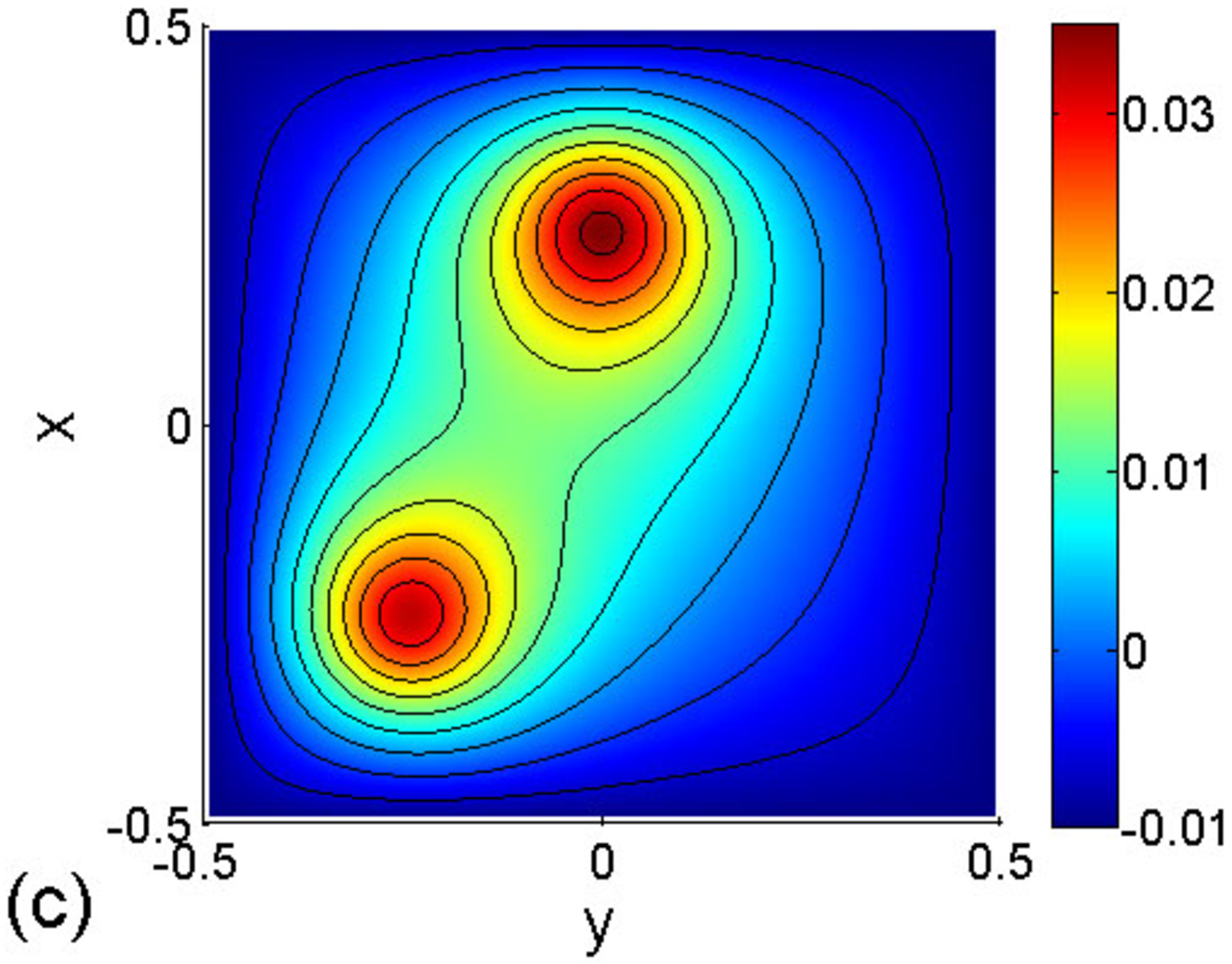}
        \includegraphics[width=4.2cm]{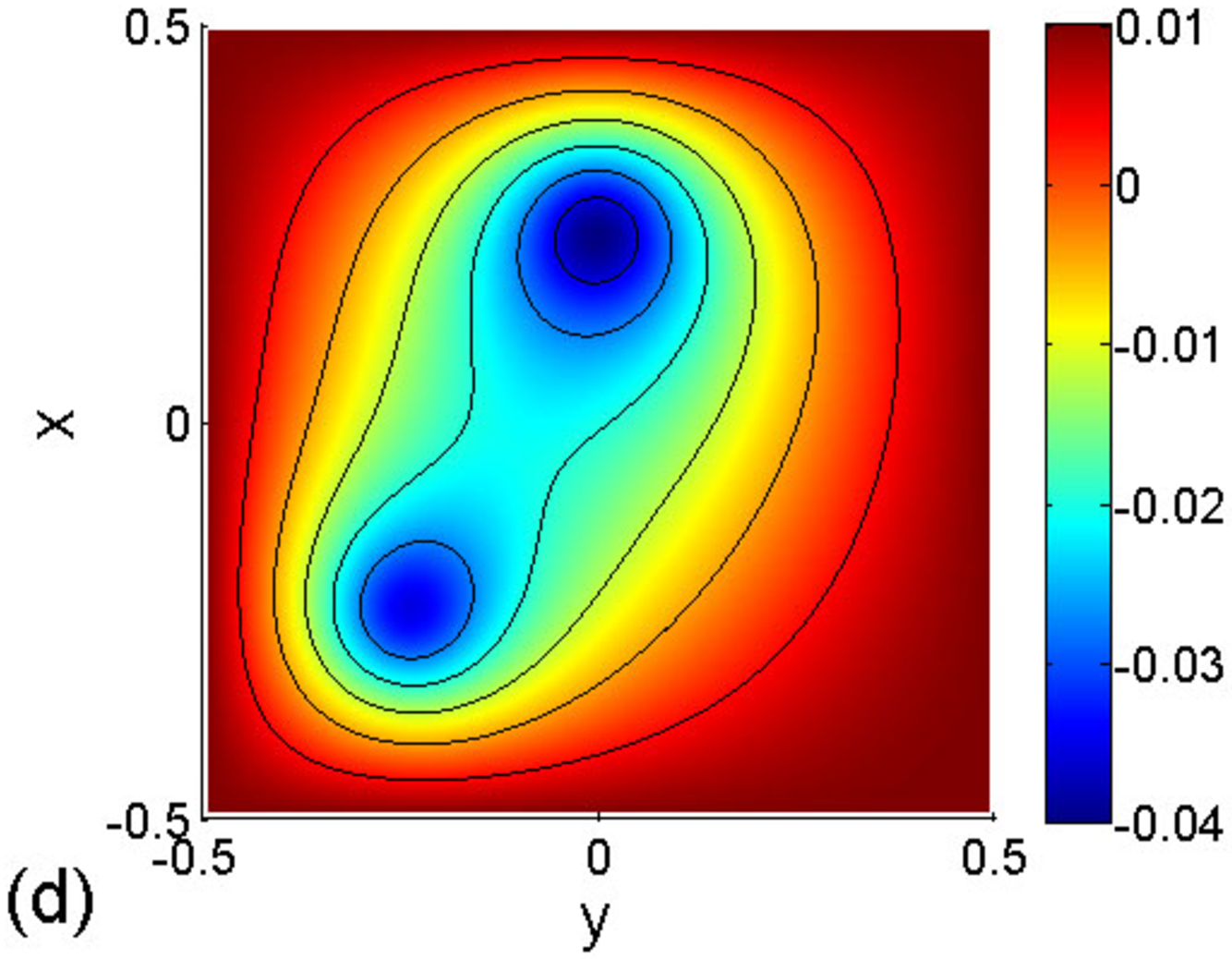}
	\caption{Seamounts: (a) topography $h(x,y)$; (b) to (d)  Equilibrium PV field for (b) $\beta=1000$, (c) $\beta=10$, (d) $\beta=-10$.}
    \label{fig:seam_b}
\end{figure}

\begin{figure}[t]
    \centering
        \includegraphics[width=4.2cm]{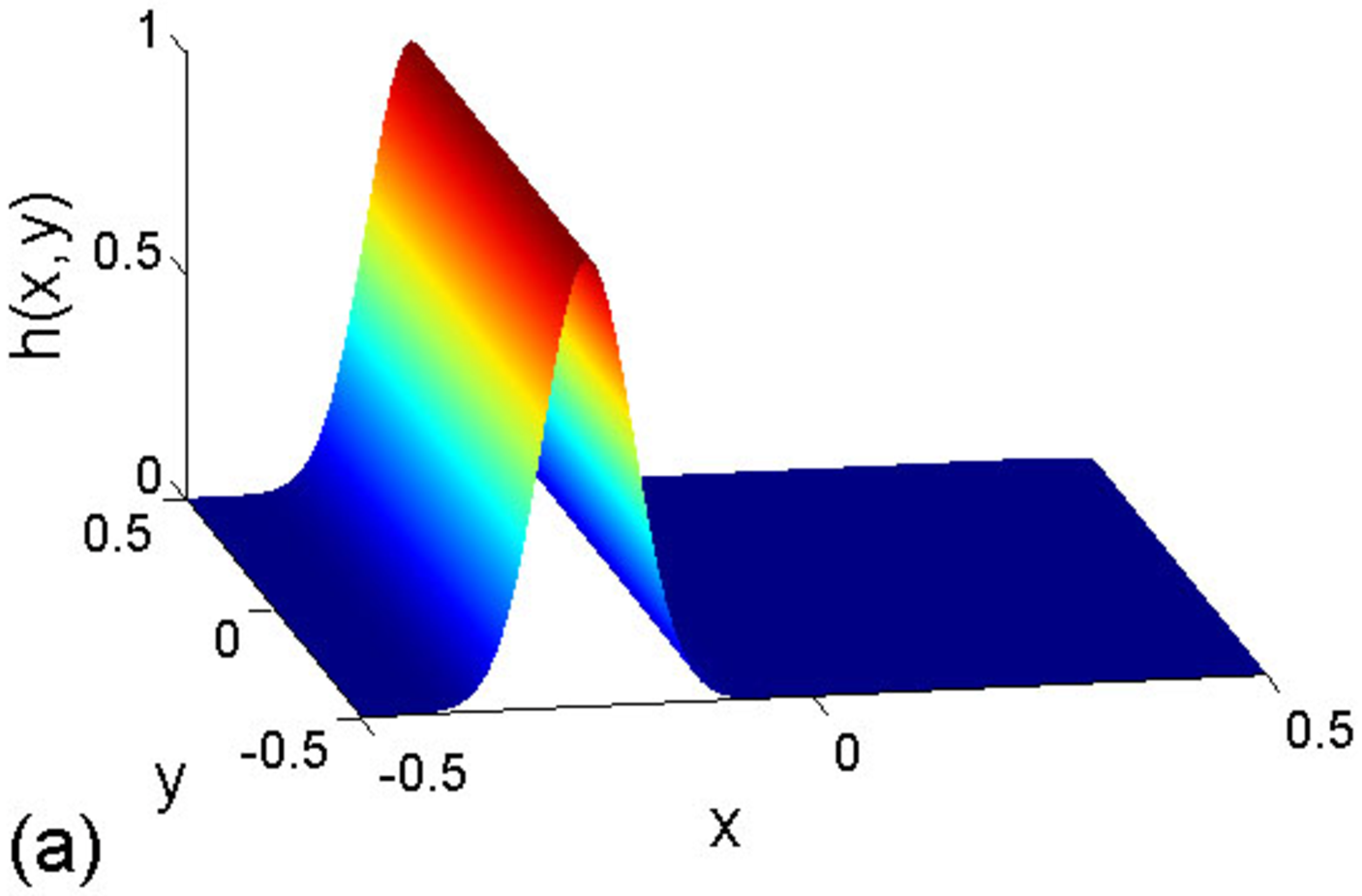}
        \includegraphics[width=4.2cm]{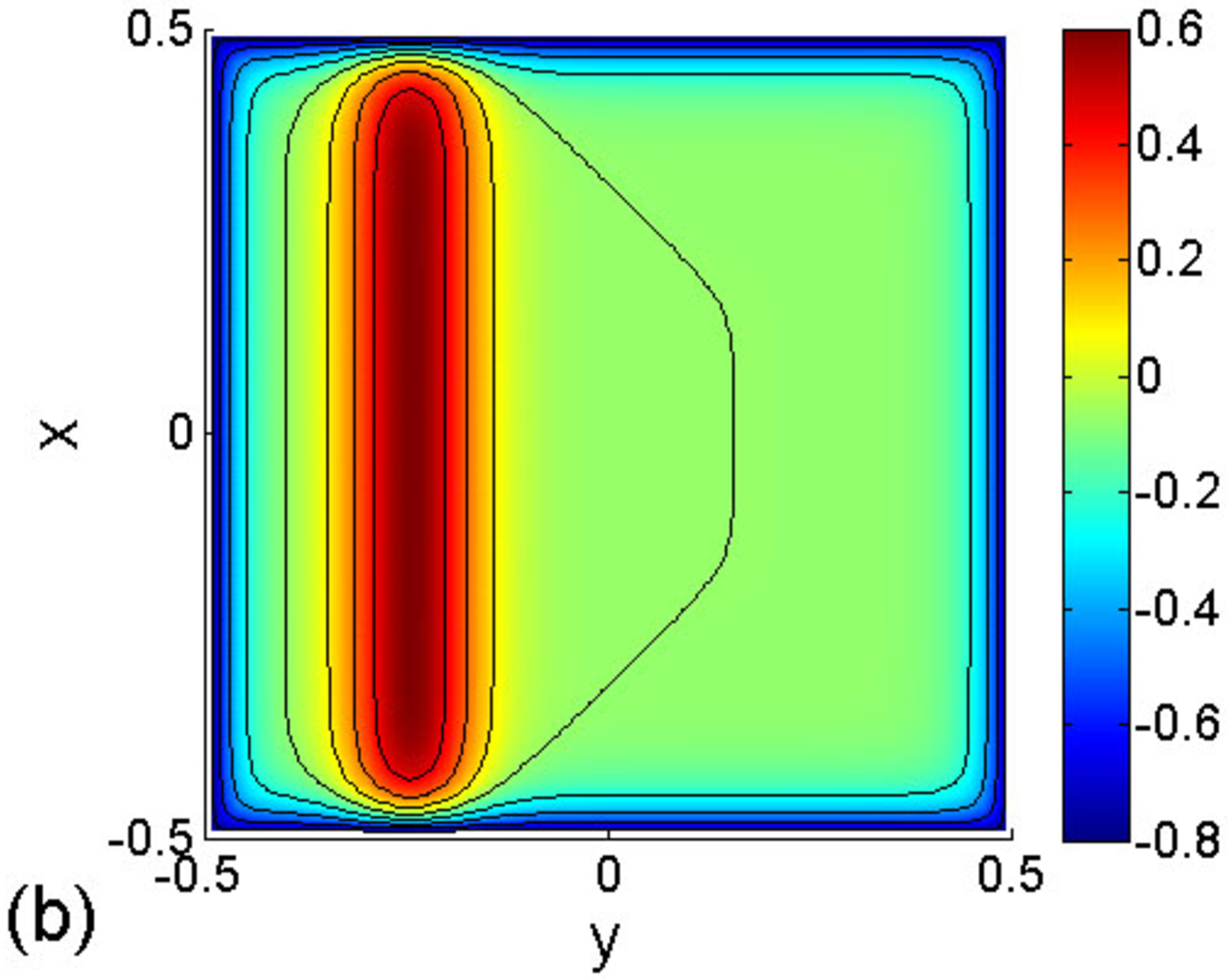}
        \includegraphics[width=4.2cm]{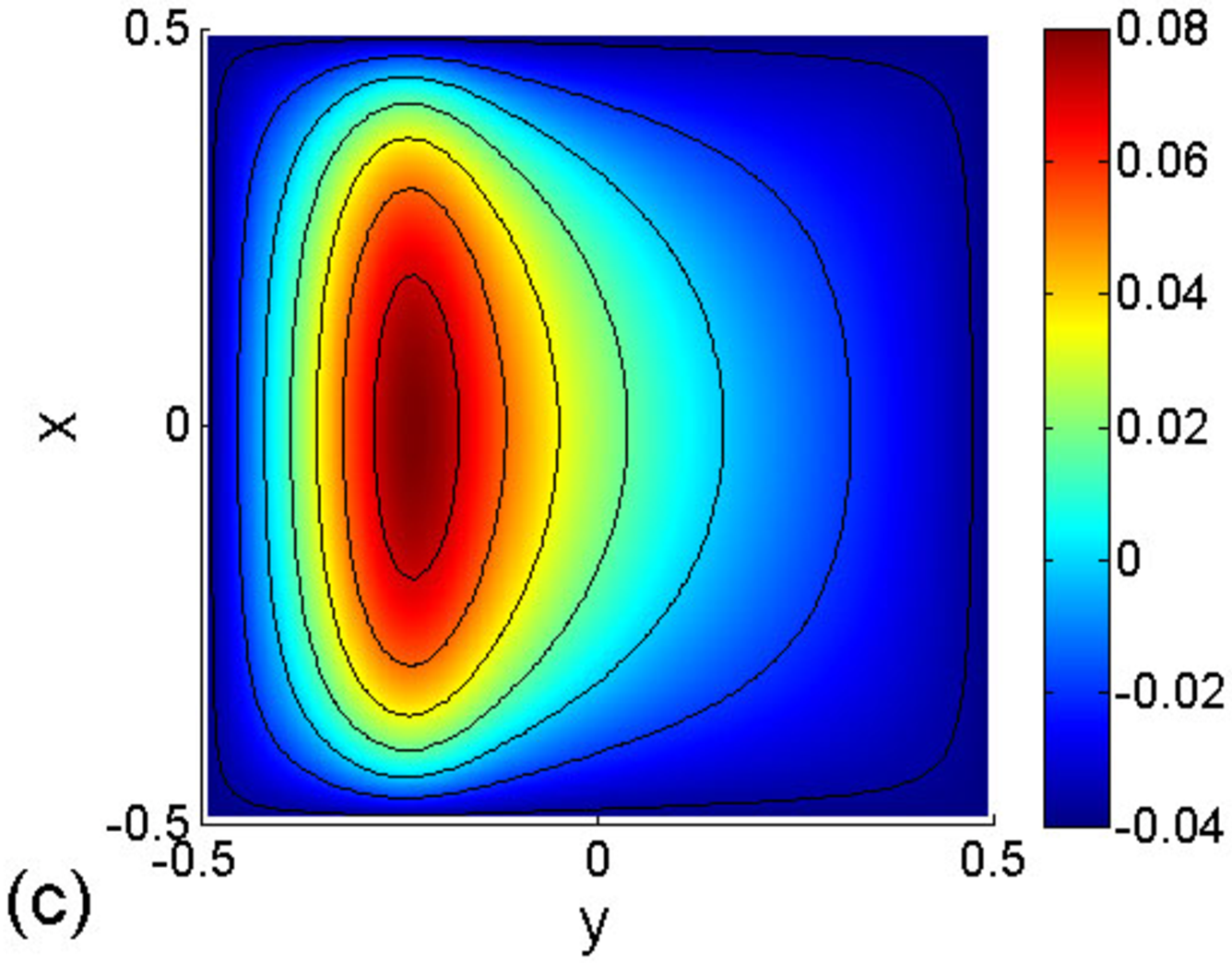}
        \includegraphics[width=4.2cm]{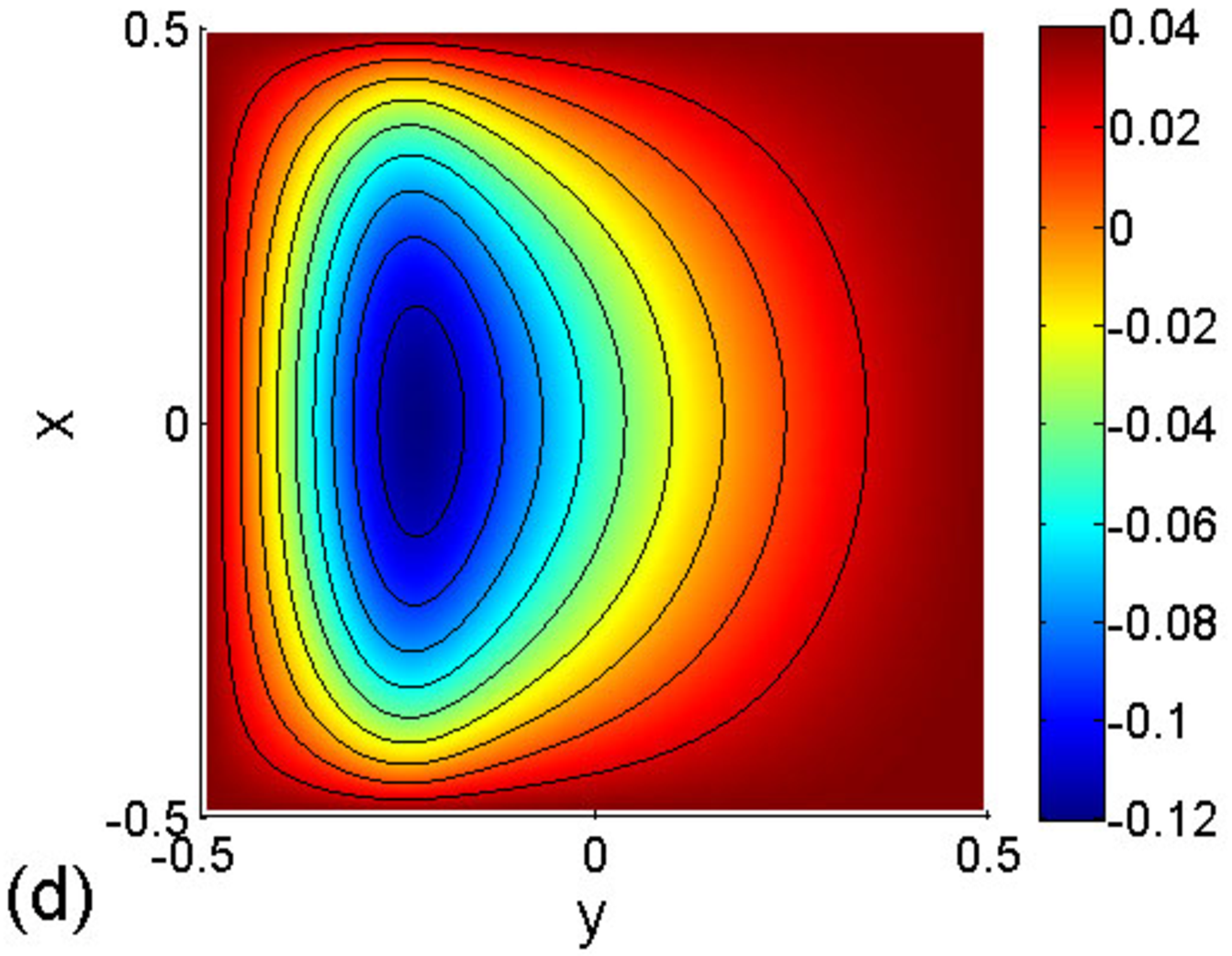}
	\caption{Ridge: (a) topography $h(x,y)$; (b) to (d) Equilibrium PV field for (b) $\beta=1000$, (c) $\beta=10$, (d) $\beta=-10$.}
    \label{fig:ridge_b}
\end{figure}

In this section, we generalize the relaxation equations derived in Sec. \ref{subs:relax_eq} so as to account for the presence of a bottom topography $h(x,y)$ and we solve them numerically to illustrate the nature of phase transitions in geophysical flows.

\subsection{The quasigeostrophic equations}

Let us consider a 2D incompressible flow over a topography $h(x,y)$ in the limit of infinite Rossby radius $R\rightarrow +\infty$. It is described by the quasigeostrophic (QG) equations
\begin{equation}
{\partial q\over\partial t}+{\bf u}\cdot \nabla q=0,\quad q=-\Delta\psi+h,\quad {\bf u}=-{\bf z}\times\nabla\psi, \label{q1}
\end{equation}
where $q$ is the potential vorticity and $\omega {\bf
z}=\nabla\times{\bf u}$ is the vorticity satisfying
$\omega=-\Delta\psi$. The QG equations conserve the energy
\begin{equation}
E=\frac{1}{2}\int(q-h)\psi d{\bf r}, \label{q2}
\end{equation}
and the Casimirs
\begin{equation}
I_f=\int f(q) d{\bf r}, \label{q3}
\end{equation}
where $f$ is an arbitrary function. We consider a generalized entropy of the form
\begin{equation}
S=-\frac{1}{2Q_2}\int \overline{q}^2 \, d{\bf r}, \label{q3b}
\end{equation}
corresponding to a Gaussian prior (see Sec. \ref{sec_gp}). As shown
in \cite{Fof}, the critical points of entropy (\ref{q3b}) at fixed
energy (\ref{q2}) and potential circulation $\Gamma=\int \overline{q}
d{\bf r}$ are solutions of the differential equation
\begin{equation}
-\Delta\psi+\beta\psi=\Gamma+\beta \langle \psi \rangle -h, \label{q4}
\end{equation}
with $\psi=0$ on the domain boundary.

We thereafter illustrate the fact that the relaxation equations
derived in Sec. \ref{subs:relax_eq} can be used as
numerical algorithms to determine maximum entropy
states. We consider three topographies $h(x,y)$ similar to
those introduced by Wang \& Vallis
\cite{WangVallis}: continental slope (see
Fig.~\ref{fig:cont_slope_b}(a)), seamounts (Fig.~\ref{fig:seam_b}(a))
and ridge (Fig.~\ref{fig:ridge_b}(a)).

\subsection{Equilibrium states}

\begin{figure}[b]
    \centering
        \includegraphics[width=9cm]{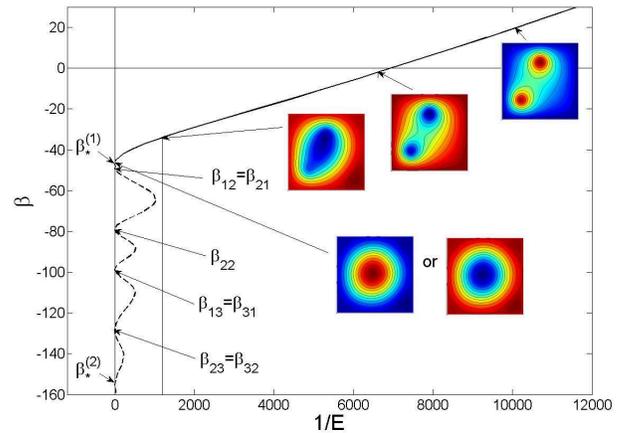}
	\caption{Relationship between $\beta$  and $1/E$ in a square domain with the topography of Fig.~\ref{fig:seam_b}(a) and $\Gamma=0$. The $q$ density is plotted for several values of $\beta$ (increasing values from blue to red).}
    \label{fig:seam_E_b}
\end{figure}

\begin{figure}[h]
    \centering
        \includegraphics[width=8cm]{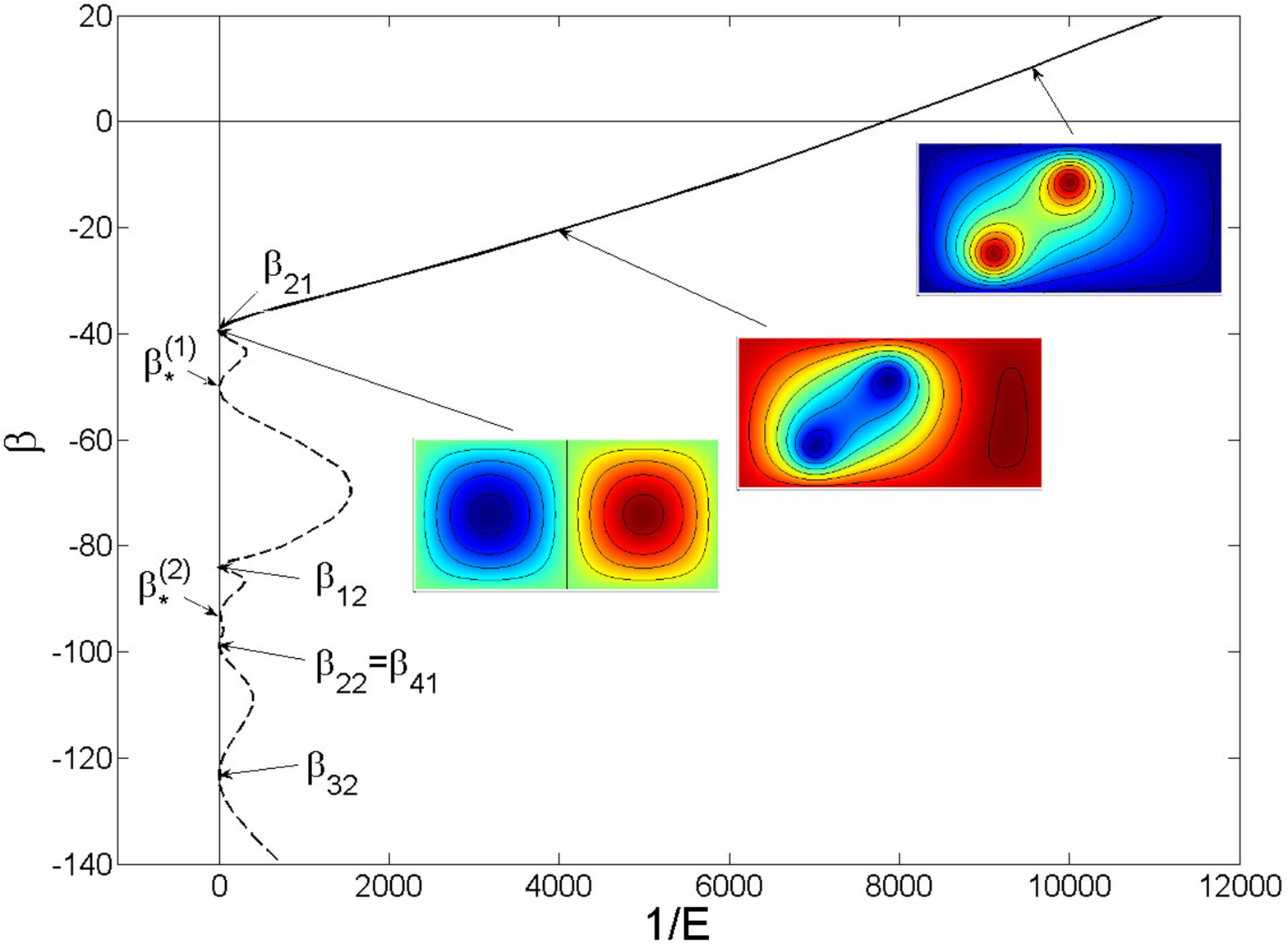}
	\caption{Relationship between $\beta$  and $1/E$ in a rectangular domain with the topography of Fig.~\ref{fig:seam_b}(a) and $\Gamma=0$ ($\tau=2$).}
    \label{fig:seam_E_b_tau_2}
\end{figure}

\begin{figure}[h]
    \centering \includegraphics[width=8cm]{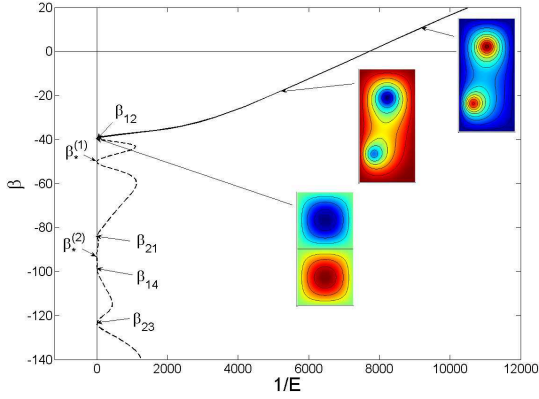}
    \caption{Relationship between  $\beta$  and $1/E$ 
    in a rectangular domain with the topography of
    Fig.~\ref{fig:seam_b}(a) and $\Gamma=0$ ($\tau=1/2$).}
    \label{fig:seam_E_b_tau_1s2}
\end{figure}

As a first step, we show in Figs.~\ref{fig:cont_slope_b},
\ref{fig:seam_b} and \ref{fig:ridge_b} the coarse-grained potential
vorticity at equilibrium for $\Gamma=0$ and different values of
$\beta$ in a square domain. These
vorticity fields have been obtained by solving
Eq.~(\ref{q4}) with the method given in
\cite{Fof}. As pointed out later, all these states
are stable (they correspond to maximum entropy states for the
corresponding values of the energy). Two
remarkable trends can be observed: (i) for very large values of
$\beta$, the potential vorticity has the tendency to
align with the topography. This corresponds to generalized Fofonoff
flows ($\psi\simeq -h/\beta$, $\overline{q}\simeq h$, ${\bf u}\simeq
\frac{1}{\beta}{\bf z}\times \nabla h$) \cite{fofonoff}; (ii) the PV fields for $\beta> 0$ and $\beta< 0$ are of opposite signs.

More quantitatively, we plot in Fig.~\ref{fig:seam_E_b} the curve
$\beta(1/E)$\footnote{Since interesting phase
transitions occur for large energies, it appears more relevant to plot
$\beta(1/E)$ instead of the more conventional caloric curve
$\beta(E)$.} for the ``seamounts'' topography in a
square domain with $\Gamma=0$. For small energies, Eq.~(\ref{q4})
admits only one solution, whereas for larger values of $E$ it admits
an infinite number of solutions, i.e. there exists an
infinite number of critical points of entropy at fixed energy and
circulation. In the latter case, the maximum entropy
state is the one with the highest $\beta$ \cite{jfm1,vb,Fof}. For
small energies, we have Fofonoff flows ($\beta>0$), for intermediate
energies we have reversed Fofonoff flows ($\beta<0$), for large
energies we have a direct monopole and for $E\rightarrow +\infty$ we
have a monopole rotating in either direction with inverse temperature
$\beta_{*}^{(1)}$. A detailed description of this caloric curve can be
found in \cite{vb,Fof}. Since $\Gamma\neq \Gamma_{*}$ (where $\Gamma_*$ is defined in \cite{Fof}), there is no
plateau at $\beta_{*}^{(1)}$. Therefore, the monopole is obtained
smoothly from the Fofonoff flows in the limit $E\rightarrow
+\infty$. In particular, in the present situation, the caloric curve $\beta(E)$ does not display
a second order phase transition.

On the other hand, Chavanis \& Sommeria \cite{jfm1} have demonstrated that 2D Euler flows characterized by a linear $\overline{\omega}-\psi$ relationship experience geometry induced phase transitions between a monopole and a dipole when the domain becomes sufficiently elongated. In the case of a rectangular domain, this transition occurs at a critical aspect ratio $\tau_{c}=1.12$ (the maximum entropy state is the monopole for $1/\tau_c<\tau<\tau_c$ and the dipole for $\tau>\tau_c$ or $\tau<1/\tau_c$). As shown by Venaille \& Bouchet \cite{vb} and Naso {\it et al.} \cite{Fof}, this property persists in the case of QG flows with a topography. To illustrate this result, we plot in Figs. \ref{fig:seam_E_b_tau_2} and \ref{fig:seam_E_b_tau_1s2} the curve
$\beta(1/E)$ for the ``seamounts'' topography in a
rectangular domain with $\Gamma=0$. We have considered rectangular
domains with aspect ratios $\tau=2>\tau_c$ (horizontal) and
$\tau=1/2<1/\tau_c$ (vertical) in order to emphasize the difference
with respect to the case of a square domain ($1/\tau_c<\tau=1<\tau_c$). For
small energies, we have Fofonoff flows ($\beta>0$), for intermediate
energies we have reversed Fofonoff flows ($\beta<0$), for large
energies we have a direct dipole and for $E\rightarrow +\infty$ we
have a dipole rotating in either direction with inverse temperature
$\beta_{21}$ (horizontal) or $\beta_{12}$ (vertical). A detailed
description of these caloric curves can be found in
\cite{vb,Fof}. Since the eigenmodes $(2,1)$ and $(1,2)$ are not
orthogonal to the topography, there is no plateau at $\beta_{21}$ or
$\beta_{12}$. Therefore, the dipole is obtained smoothly from the
Fofonoff flows in the limit $E\rightarrow +\infty$. In particular, in the present situation, the
caloric curve $\beta(E)$ does not display a second order phase
transition.

As a final remark, we note that low energy states are strongly
influenced by the topography (and only weakly by the domain geometry)
as in the Fofonoff \cite{fofonoff} study whereas high energy states are
strongly influenced by the domain geometry (and only weakly by the
topography) as in the Chavanis-Sommeria \cite{jfm1} study.

\subsection{Relaxation towards the maximum entropy state}

Following the same approach as in Sec. \ref{sec_pn} and considering a Gaussian prior like in Sec. \ref{subs:relax_eq}, one can derive the following relaxation equations for the coarse-grained potential vorticity and its local centered variance
\begin{eqnarray}
\frac{\partial \overline{q}}{\partial t}+{\bf u}\cdot\nabla\overline{q}=-D\left[ \frac{\overline{q}}{Q_2}+\beta(t)\psi+\alpha(t) \right], \label{q5} \\
\frac{\partial q_2}{\partial t}+{\bf u}\cdot\nabla q_2=2D \left( 1-\frac{q_2}{Q_2} \right). \label{q6}
\end{eqnarray}
The conservation of $E$ and $\Gamma$ implies
\begin{eqnarray}
\beta(t)=\frac{1}{Q_2}\frac{\Gamma\langle\psi\rangle-A (2E+ \langle \psi h \rangle)}{A\langle\psi^2\rangle-\langle\psi\rangle^2}, \label{q7} \\
\alpha(t)=-\frac{1}{Q_2}\frac{\Gamma\langle\psi^2\rangle-\langle\psi\rangle(2E+ \langle \psi h \rangle)}{A\langle\psi^2\rangle-\langle\psi\rangle^2}. \label{q8}
\end{eqnarray}
We thereafter integrate numerically Eqs.~(\ref{q5})-(\ref{q8}) with the following boundary conditions
\begin{eqnarray}
\psi|_{\partial D}=0,\\
\overline{q}|_{\partial D}=-Q_2\alpha(t),\\
q_2|_{\partial D}=Q_2,
\end{eqnarray}
where $\partial D$ is the domain boundary.  In the
simulations, all the quantities are normalized by the lengthscale
$A^{1/2}$ and by the timescale $Q_2^{-1/2}$ (this
amounts to taking $A=Q_2=1$ in the foregoing equations). Furthermore,
the potential circulation $\Gamma$ is taken equal to $0$ and the energy is normalized by $2/b^2$.

\begin{figure}[h]
    \centering
        \includegraphics[width=8cm]{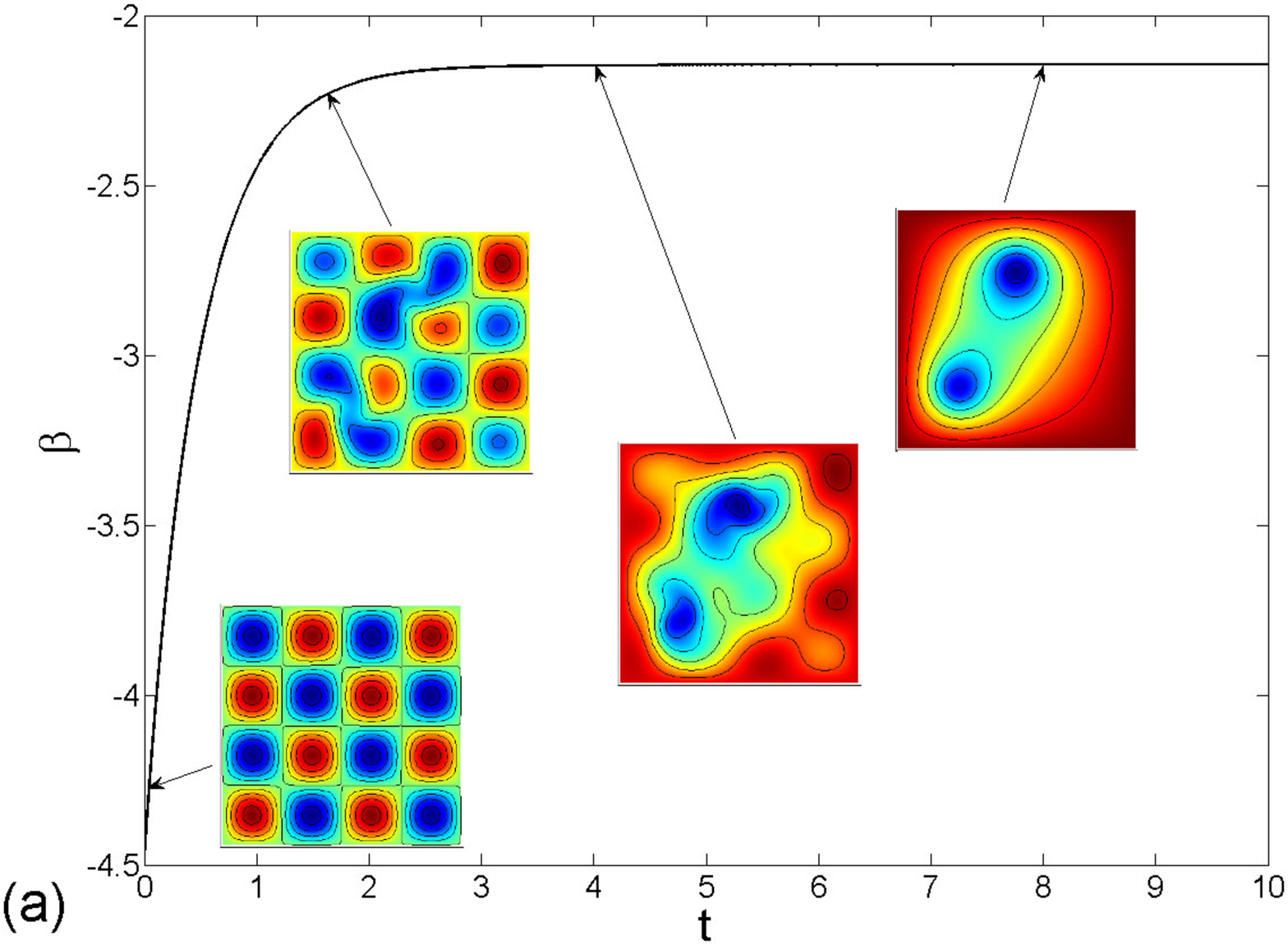}\\
        \includegraphics[width=8cm]{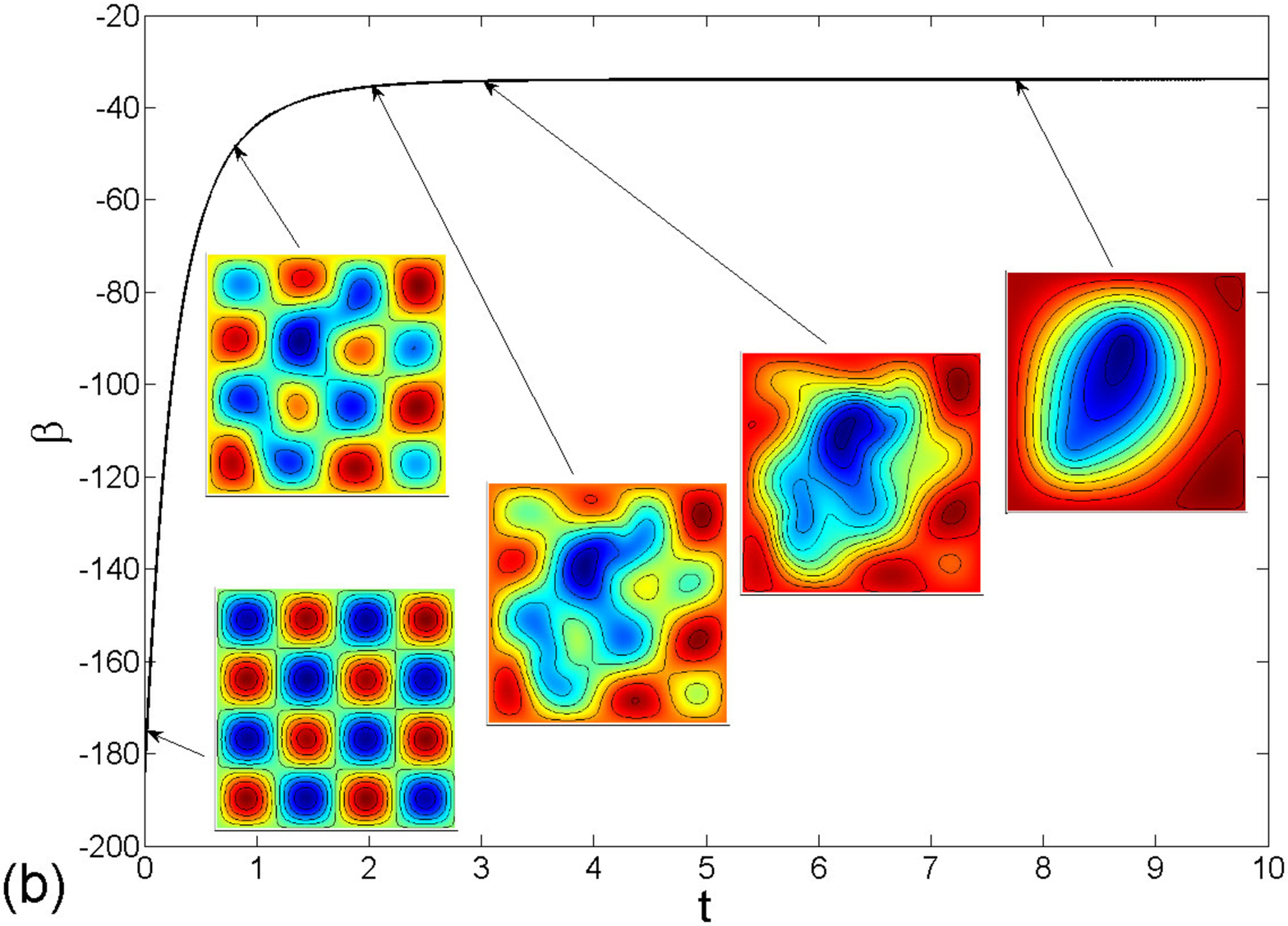}\\
        \includegraphics[width=8cm]{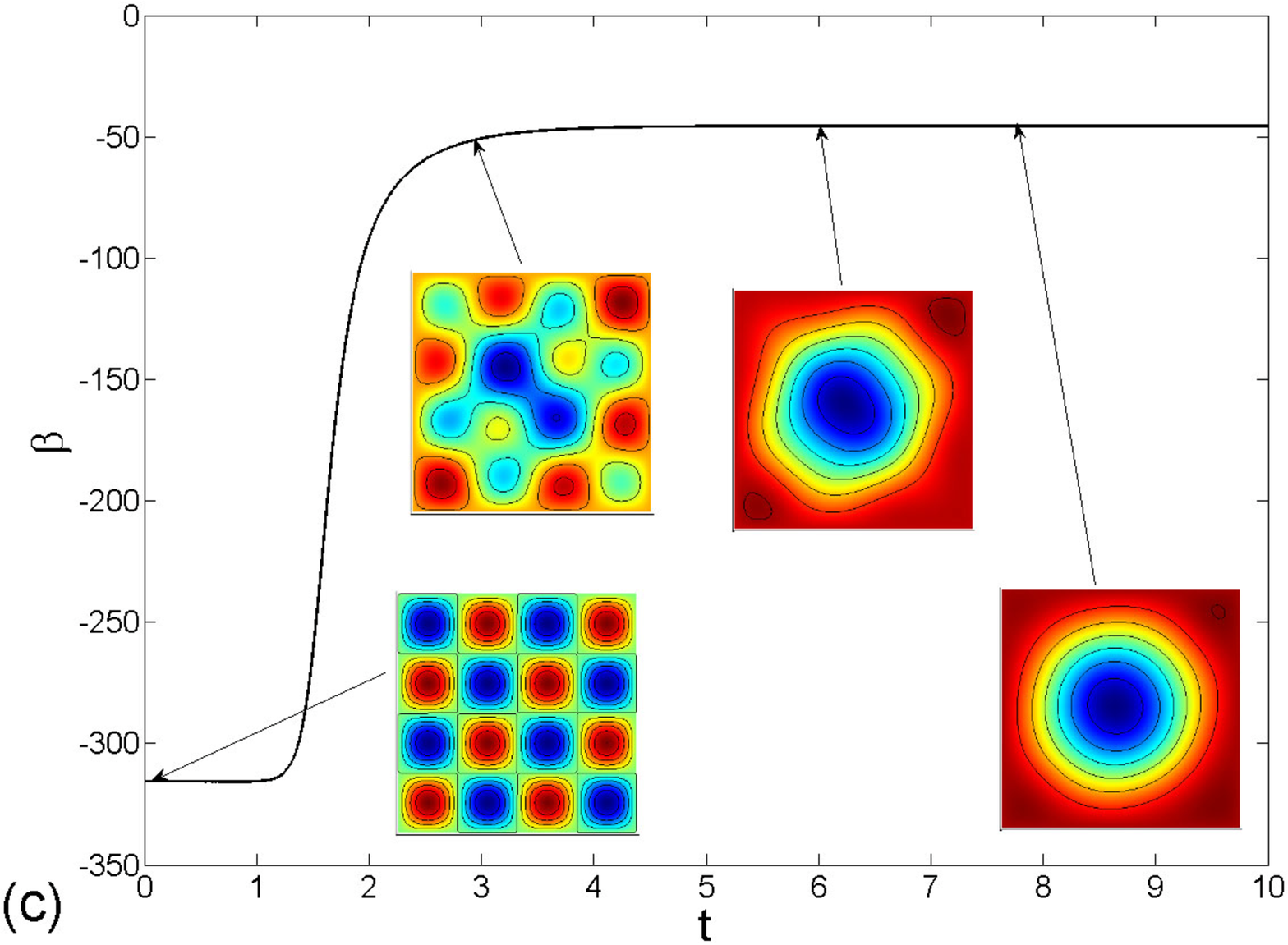}
	\caption{Relaxation of the system at fixed $E$, with the topography of Fig.~\ref{fig:seam_b}(a), $D=1$ and $\Gamma=0$ (square domain): (a) $1/E\approx 6620$, $\beta_f\approx -2.1$ (see vertical line in Fig.~\ref{fig:seam_E_b}), (b) $1/E\approx 1196$, $\beta_f\approx -34$ (see vertical line in Fig.~\ref{fig:seam_E_b}), (c) $1/E\approx 1.3.10^{-5}$, $\beta_f\approx \beta_*^{(1)}$.}
    \label{fig:seam_relax_E}
\end{figure}

We first illustrate the convergence of these relaxation equations
towards the maximum entropy state, integrating them
with initial condition:
$\overline{q}(x,y,t=0)=q_0\sin(4\pi x)\sin(4\pi y)$ and
$q_2(x,y,t=0)=\sin^2(2\pi x)\sin^2(2\pi y)+Q_2$. Varying the value of
the parameter $q_0$ enables us to generate initial conditions of
different energy $E$. The dynamics of $\overline{q}$
and $q_2$ are not coupled (except through the advective term). We therefore do not show here the time
evolution of $q_2$ that always converges to the uniform state
$q_2(x,y)=Q_2$.

\begin{figure}[h]
    \centering
        \includegraphics[width=8cm]{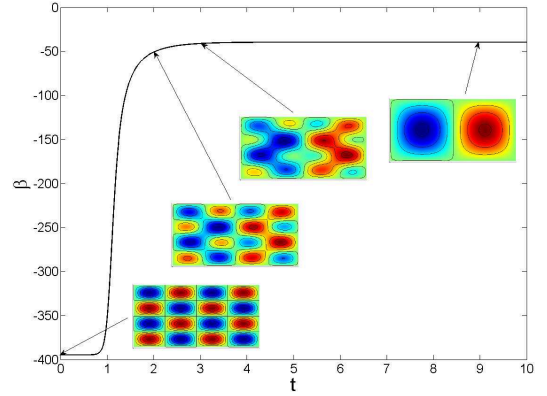}
	\caption{Relaxation of the system at fixed $E$, with the topography of Fig.~\ref{fig:seam_b}, $D=1$, $\Gamma=0$, $\tau=2$ and $1/E\approx 1.6.10^{-5}$, $\beta_f\approx \beta_{21}$ (see Fig.~\ref{fig:seam_E_b_tau_2}).}
    \label{fig:seam_relax_tau_2}
\end{figure}

We plot in Fig.~\ref{fig:seam_relax_E}
the time evolution of
$\beta(t)$ and  $\overline{q}({\bf r},t)$
 for three values of $E$ in a
square domain. For each value of energy $E$, the
relaxation equations converge towards the corresponding maximum
entropy state (see plateaus in Fig.~\ref{fig:seam_relax_E}). The
value of $\beta$ in the final state, thereafter denoted $\beta_f$,
is in every case in agreement with the value reported
in Fig.~\ref{fig:seam_E_b}. For large values of $1/E$
(Fig.~\ref{fig:seam_relax_E}(a) and (b)), the
relaxation equations converge towards a Fofonoff flow which is their
unique steady state. A case of interest is the one arising for
$1/E=0$. In that case, there exists an infinite
number of steady states and the initial condition that we have imposed
is an unstable steady state. After some time, the system
destabilizes and finally converges towards the (stable) maximum
entropy state. In a square domain, the maximum entropy state of
infinite energy is the monopole for which $\beta=\beta_*^{(1)}$. The
limit of infinite energy cannot be reached numerically, but we can
approach it (see Fig.~\ref{fig:seam_relax_E}(c)). As expected, for
very large but finite values of $E$, the stable steady state is
the \textit{direct} monopole which is the natural
evolution of a Fofonoff flow as energy increases. We plot in
Fig.~\ref{fig:seam_relax_tau_2} the time evolution of $\beta(t)$ and
$\overline{q}({\bf r},t)$ in a rectangular domain with aspect ratio
$\tau=2>\tau_c$. For a sufficiently large energy, the system evolves
towards a dipole instead of a monopole.

\begin{figure}[h]
    \centering
        \includegraphics[width=8cm]{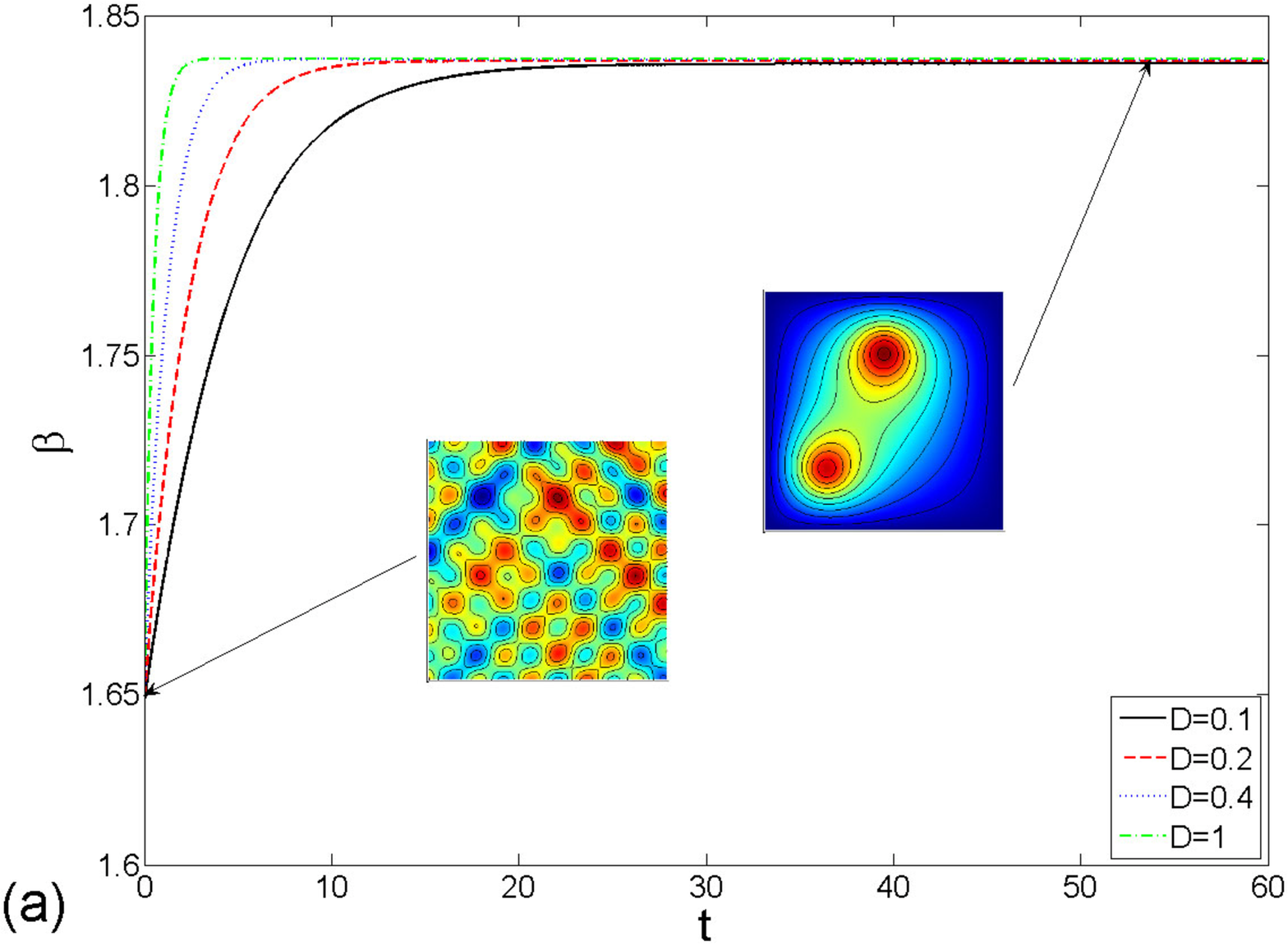}\\
        \includegraphics[width=8cm]{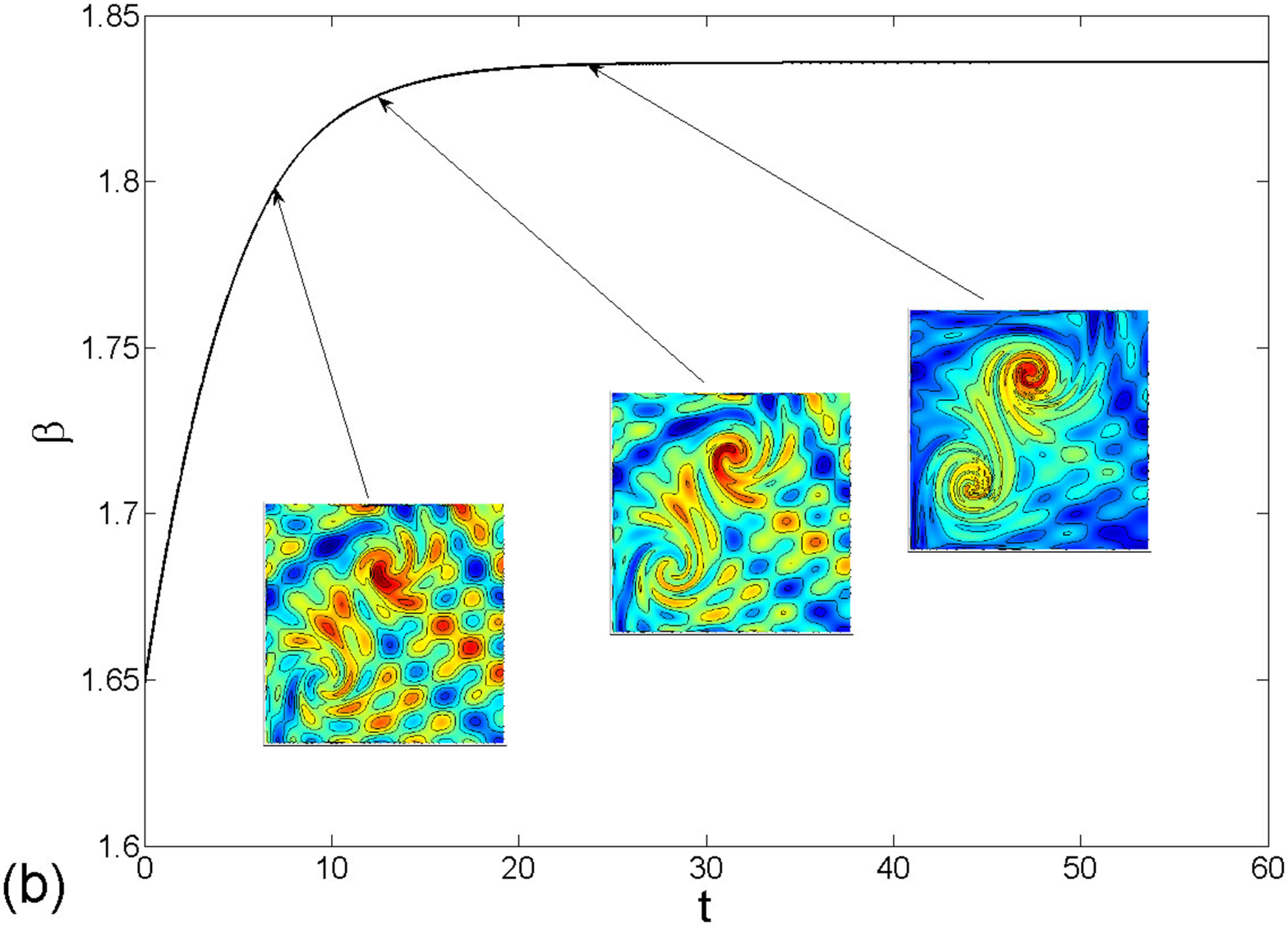}\\
		\includegraphics[width=8cm]{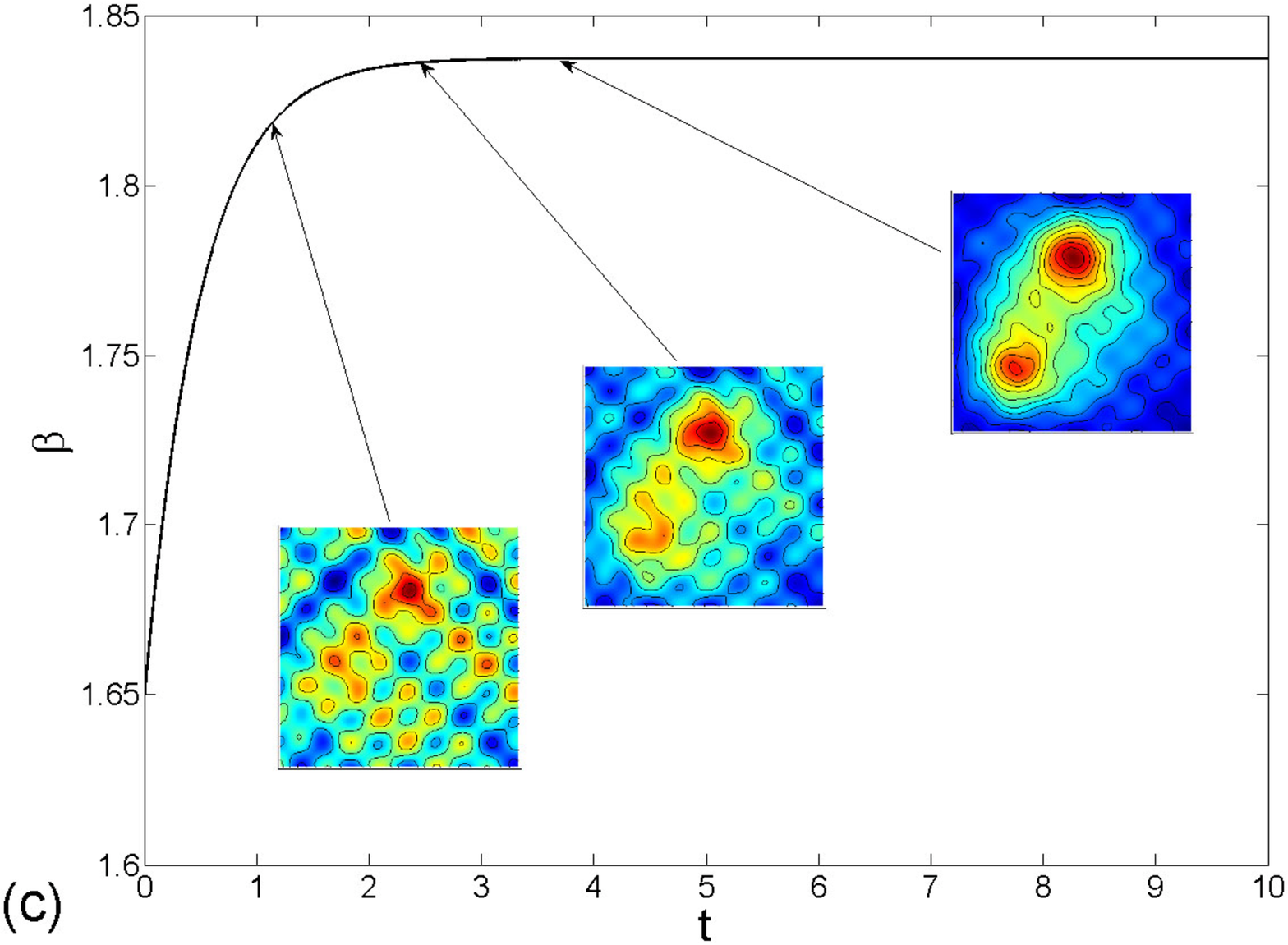}
	\caption{Relaxation of the system at fixed $E$, with the topography of Fig.~\ref{fig:seam_b}(a) and $\Gamma=0$ (square domain): (a) $D=0.1,0.2,0.4,1$; (b) $D=0.1$; (c) $D=1$. For small values of $D$, the mixing is efficient and leads to filament-like structures in the transient states.}
    \label{fig:seam_relax_D}
\end{figure}

We now investigate the influence of the relaxation term (r.h.s. in
Eq. (\ref{q5})) on the dynamics. To that purpose, we integrate the
relaxation equations with different values of $D$, starting from the
same initial condition: a random field written as the sum of sine
functions with random phases and amplitudes, and wave numbers ranging
from $1$ to $5$. As shown in Fig.~\ref{fig:seam_relax_D}(a),
corresponding to an energy $1/E=7200$, the system always relaxes
towards the maximum entropy state, with $\beta_f\approx 1.83$. As
expected, this state is reached at longer times when $D$ is
decreased. More interestingly, $D$ measures the relative importance of
the relaxation term (that forces the convergence towards the maximum
entropy state) and of the advection (related to the
mixing). Therefore, the intermediate vorticity fields are not
identical for different values of $D$. For a smaller $D$, the
advection can play a larger role (i.e. the mixing is more efficient)
before the final state is reached. This behavior is clearly seen by
comparing Figs.~\ref{fig:seam_relax_D}(b) and (c). If
we use the relaxation equations as a parametrization of 2D turbulence
(see, however, the last paragraph of the conclusion), we see that
the influence of the relaxation term is to smooth out the small scales
without strong influence on the large-scale dynamics. This is
precisely the role of a parametrization since we are in general
interested by the largest scales, not by the fine structure of the
flow. Our proposed parametrization rigorously conserves the
circulation and the energy (contrary to a parametrization involving
only a turbulent viscosity or an hyperviscosity) and ``pushes'' the
system towards the statistical equilibrium state with Gaussian
fluctuations. Therefore, it allows to simulate 2D flows with less
resolution than usually required (the simulations have been performed
on a $128^2$ grid) since the small scales have been modelled in an
optimal manner. It would be interesting, however, to compare the
efficiency of different parametrizations.

\section{Conclusion} \label{sec_conclusion}

In this paper, we have proposed a new class of relaxation equations
associated with the Ellis-Haven-Turkington statistical theory of 2D
turbulence where a prior vorticity distribution is prescribed instead
of the Casimir constraints in the Miller-Robert-Sommeria theory. We
have considered specifically the case of a Gaussian prior associated
with minimum enstrophy states and we have given a numerical
illustration of these relaxation equations in connection to Fofonoff
flows in oceanic circulation.  We have discussed the connections with
previous relaxation equations introduced by Robert \& Sommeria
\cite{rsmepp} and Chavanis \cite{gen,physicaD,aussois}.  These relaxation
equations can provide efficient numerical algorithms to solve the
various constrained maximization problems appearing in the statistical
mechanics of 2D turbulence \cite{proc}. This is clearly an interest of
these equations because it is usually difficult to directly solve the
Euler-Lagrange equations (\ref{mrs7}) for the critical points of
entropy and be sure that they are entropy maxima. The relaxation
equations assure, by construction, that the relaxed state is a maximum
entropy state with appropriate constraints. In this sense, the
relaxation equations can provide an alternative and complementary
method to the numerical algorithm of Turkington \& Whitaker \cite{tw}
that has been introduced to solve constrained optimization
problems. An advantage of using relaxation equations
is that, by incorporating a space and time dependent diffusivity
related to the strength of the fluctuations
\cite{rr,csr}, one can heuristically account for {\it incomplete relaxation} \cite{jfm2,brands},
which is not possible with the Turkington-Whitaker
algorithm. Furthermore, even if these relaxation equations cannot be
considered as a parametrization of 2D turbulence (see below), they may
nevertheless provide an idea of the true dynamics towards
equilibrium. In that respect, it would be interesting to compare them
with large eddy simulations (LES).  These relaxation equations also
provide new classes of partial differential equations which can be of
interest to mathematicians. In fact, several mathematical works have
started to study these classes of relaxation equations
\cite{rm,ros1,ros2}. The connection to nonlinear mean field
Fokker-Planck equations and to the Keller-Segel model of chemotaxis in
biology is also interesting to mention in that respect \cite{nfp}. If
we want to address more fundamental issues from first principles,
we can try to develop kinetic theories such as the quasilinear theory
of the 2D Euler equation \cite{quasi}, the rapid distortion theory
\cite{dn} or the stochastic structural stability theory
\cite{fi}. In that respect, we would like to briefly
comment on the $H$-theorem in 2D turbulence. Ideally, the kinetic
equations for $\rho({\bf r},\sigma,t)$ and $\overline{\omega}({\bf
r},t)$ should be derived from first principles, as attempted in
\cite{quasi}, and the $H$-theorem should be derived from these kinetic
equations. Here, we have used the inverse approach: assuming that an
$H$-theorem holds for some functionals, we have constructed relaxation
equations that satisfy this $H$-theorem while conserving some
constraints. This is clearly a purely {\it ad hoc} procedure that is
sufficient to construct numerical algorithms solving constrained
maximization problems, but not sufficient to conclude that these
relaxation equations constitute an accurate parametrization of 2D
turbulence. Our main goal, here and in \cite{proc}, was to provide a
relaxation equation for each optimization problem introduced in 2D
turbulence. Therefore, there exists as many relaxation equations as
maximization problems. These relaxation equations can be used as
numerical algorithms to solve these optimization problems and
therefore determine dynamically and/or thermodynamically stable steady
states. This is the main virtue of these relaxation equations
\cite{proc}.

\appendix

\section{Maximization of entropy in three steps} \label{sec_meh}

Let us consider the maximization problem (see also \cite{minens}):
\begin{eqnarray}
\label{meh1}
\max_{\rho}\quad \lbrace S[\rho]\quad |\quad \Gamma[\overline{\omega}]=\Gamma, \quad E[\overline{\omega}]=E, \nonumber\\ \Gamma_2^{f.g}[\overline{\omega^2}]=\Gamma_2^{f.g}, \quad \int\rho\, d\sigma=1 \rbrace,
\end{eqnarray}
where $S$ is the MRS entropy (\ref{mrs1}). To solve the maximization problem (\ref{meh1}), we can proceed in three steps. This will show the connection with the results of Sec. \ref{sec_moment}.

{\it First step:} we first maximize $S[\rho]$ at fixed $E$, $\Gamma$, $\Gamma_2^{f.g.}$, normalization {\it and} a fixed profile of vorticity $\overline{\omega}({\bf r})=\int\rho\sigma\, d\sigma$ and enstrophy   $\overline{\omega^2}({\bf r})=\int\rho\sigma^2\, d\sigma$. Since the specification of $\overline{\omega}({\bf r})$ and  $\overline{\omega^2}({\bf r})$
determines $E$, $\Gamma$ and $\Gamma_2^{f.g.}$, this is equivalent to maximizing $S[\rho]$ at fixed normalization for a fixed profile of vorticity $\overline{\omega}({\bf r})=\int\rho\sigma\, d\sigma$ and enstrophy   $\overline{\omega^2}({\bf r})=\int\rho\sigma^2\, d\sigma$. The global entropy maximum of this problem is the distribution (\ref{gauss}). Then, we can express the entropy  (\ref{mrs1}) in terms of $\overline{\omega}$ and $\overline{\omega^2}$ by substituting the optimal distribution  (\ref{gauss}) in Eq. (\ref{mrs1}). This gives the functional (\ref{entom}).

{\it Second step:} The maximization problem (\ref{meh1}) is now equivalent to
\begin{eqnarray}
\label{meh2}
\max_{\overline{\omega},\overline{\omega^2}}\quad \lbrace S[\overline{\omega},\overline{\omega^2}]\quad |\quad \Gamma[\overline{\omega}]=\Gamma, \quad E[\overline{\omega}]=E, \nonumber\\ \Gamma_2^{f.g}[\overline{\omega^2}]=\Gamma_2^{f.g} \rbrace,
\end{eqnarray}
where $S[\overline{\omega},\overline{\omega^2}]$ is given by
Eq. (\ref{entom}). To solve that problem, we now maximize
$S[\overline{\omega},\overline{\omega^2}]$ at fixed $E$, $\Gamma$,
$\Gamma_2^{f.g.}$ {\it and} a fixed vorticity profile 
$\overline{\omega}({\bf r})$. Since the specification of
$\overline{\omega}({\bf r})$ determines $E$, $\Gamma$ and
$\Gamma_2^{c.g.}$, this is equivalent to maximizing
$S[\overline{\omega},\overline{\omega^2}]$ at fixed 
vorticity profile $\overline{\omega}({\bf r})$ and fixed $\int\omega_2\, d{\bf
r}=\Gamma_2^{f.g.}-\Gamma_2^{c.g.}$. The global entropy maximum of
this problem is $\omega_2({\bf
r})=\Omega_2=(\Gamma_2^{f.g.}-\Gamma_2^{c.g.})/A$,
where $A$ is the domain area. Then, we can express the entropy
(\ref{entom}) in terms of $\overline{\omega}$ by substituting the
foregoing solution in Eq. (\ref{entom}). This gives the functional
\begin{eqnarray}
\label{meh3}
S[\overline{\omega}]=\frac{A}{2}\ln\Omega_2=\frac{A}{2}\ln\left (\Gamma_2^{f.g.}-\Gamma_2^{c.g.}[\overline{\omega}]\right ),
\end{eqnarray}
up to an additional constant.

{\it Third step:} The maximization problem (\ref{meh2}) is now equivalent to
\begin{eqnarray}
\label{meh4}
\max_{\overline{\omega}}\quad \lbrace S[\overline{\omega}]\quad |\quad \Gamma[\overline{\omega}]=\Gamma, \quad E[\overline{\omega}]=E\rbrace.
\end{eqnarray}
where $S[\overline{\omega}]$ is given by Eq. (\ref{meh3}). Since $\ln(x)$ is a monotonically increasing function, we can finally remark that this maximization problem is equivalent to
\begin{eqnarray}
\label{meh5}
\min_{\overline{\omega}}\quad \lbrace \Gamma_2^{c.g.}[\overline{\omega}]\quad |\quad \Gamma[\overline{\omega}]=\Gamma, \quad E[\overline{\omega}]=E\rbrace.
\end{eqnarray}
In conclusion, the maximization of entropy at fixed circulation, energy and microscopic enstrophy  (\ref{meh1}) and (\ref{meh2})  are equivalent to the minimization of coarse-grained enstrophy at fixed circulation and energy (\ref{meh5}) \cite{minens}.

{\it Remark}: Writing the variational problem  associated to (\ref{meh4})  in the form $\delta S-\beta\delta E-\alpha\delta\Gamma=0$ (first variations), we recover Eq. (\ref{ajy}).

\section{Equivalence between the stability criteria (\ref{gmrs8}) and (\ref{gmrs13})}
\label{sec_eqa}

In Sec. \ref{sec_get}, we have shown the equivalence of (\ref{gmrs3})
and (\ref{gmrs12}) for global maximization.  Another proof is given by
Ellis {\it et al.} \cite{ellis} by using large deviations technics
(they show that the probability density of the coarse-grained
vorticity $P[\overline{\omega}]$ is given by a Cramer formula
involving the entropy (\ref{genent})-(\ref{ge8}), so that the most
probable coarse-grained vorticity solves the maximization problem
(\ref{gmrs12})). In this Appendix, we show the equivalence of
(\ref{gmrs3}) and (\ref{gmrs12}) for local maximization,
i.e. $\rho({\bf r},\sigma)$ is a (local) maximum of $S_{\chi}[\rho]$
at fixed $E$, $\Gamma$ and normalization if, and only if, the
corresponding coarse-grained vorticity $\overline{\omega}({\bf r})$ is
a (local) maximum of $S[\overline{\omega}]$ at fixed $E$ and
$\Gamma$. To that purpose, using a method sketched in \cite{phd}, we
show the equivalence between the stability criteria (\ref{gmrs8}) and
(\ref{gmrs13}). Another proof is given by Bouchet \cite{bouchet} (see
also Appendix G of Chavanis
\cite{proc}) by using an orthogonal decomposition of the perturbation
\cite{frank}.

We shall determine the perturbation $\delta\rho_{*}({\bf r},\sigma)$ that maximizes $\delta^{2}J[\delta\rho]$ given by Eq. (\ref{gmrs8}) with the constraints $\delta\overline{\omega}=\int \delta\rho\sigma\, d\sigma$ and $\int \delta\rho\, d\sigma=0$, where $\delta\overline{\omega}({\bf r})$ is prescribed (assumed to conserve energy and circulation at first order). Since the specification of  $\delta\overline{\omega}$ determines $\delta\psi$, hence the second integral in Eq. (\ref{gmrs8}), we can write the variational problem in the form
\begin{eqnarray}
\label{eqa1}
\delta\left (-\frac{1}{2}\int\frac{(\delta\rho)^2}{\rho}\, d{\bf r}d\sigma\right )-\int\lambda({\bf r})\delta\left (\int\delta\rho\sigma\, d\sigma\right )\, d{\bf r}\nonumber\\
- \int\zeta({\bf r})\delta\left (\int\delta\rho\, d\sigma\right )\, d{\bf r}=0,\quad 
\end{eqnarray}
where $\lambda({\bf r})$ and $\zeta({\bf r})$ are Lagrange multipliers. This gives
\begin{eqnarray}
\label{eqa2}
\delta\rho_*=-\rho({\bf r},\sigma)(\lambda({\bf r})\sigma+\zeta({\bf r})),
\end{eqnarray}
and it is a global maximum of $\delta^{2}J[\delta\rho]$ with the previous constraints since $\delta^2(\delta^2 J)=-\int \frac{\delta(\delta\rho)^2}{2\rho}\, d{\bf r}d\sigma< 0$ (the constraints are linear in $\delta\rho$ so their second variations vanish). The Lagrange multipliers are determined from the constraints $\delta\overline{\omega}=\int \delta\rho\sigma\, d\sigma$ and $\int \delta\rho\, d\sigma=0$ yielding $\delta\overline{\omega}=-\lambda\overline{\omega^2}-\zeta\overline{\omega}$ and $0=-\lambda\overline{\omega}-\zeta$. Therefore, the optimal perturbation (\ref{eqa2}) can finally be written \begin{eqnarray}
\label{eqa3}
\delta\rho_*=\frac{\delta\overline{\omega}}{\omega_2}\rho(\sigma-\overline{\omega}).
\end{eqnarray}
Since it maximizes $\delta^{2}J[\delta\rho]$, we have
$\delta^{2}J[\delta\rho]\le \delta^{2}J[\delta\rho_*]$. Explicating
$\delta^{2}J[\delta\rho_*]$ using Eqs. (\ref{gmrs8}) and (\ref{eqa3}),
we obtain
\begin{eqnarray}
\label{eqa3b}
\delta^{2}J[\delta\rho]\le -\frac{1}{2}\int \frac{(\delta\overline{\omega})^2}{\omega_2}\, d{\bf r}-\frac{1}{2}\beta\int\delta\overline{\omega}\delta\psi\, d{\bf r}.
\end{eqnarray}
Finally, using Eq. (\ref{ind}), the foregoing inequality can be rewritten
\begin{eqnarray}
\label{eqa4}
\delta^{2}J[\delta\rho]\le -\frac{1}{2}\int C''(\overline{\omega})(\delta\overline{\omega})^2\, d{\bf r}\nonumber\\
-\frac{1}{2}\beta\int\delta\overline{\omega}\delta\psi\, d{\bf r}\equiv \delta^{2}J[\delta\overline{\omega}],
\end{eqnarray}
where the r.h.s. is precisely the functional appearing in Eq. (\ref{gmrs13}). Furthermore, there is equality in Eq. (\ref{eqa4}) iff $\delta\rho=\delta\rho_*$. This proves that the stability criteria (\ref{gmrs8}) and (\ref{gmrs13}) are equivalent. Indeed: (i) if inequality (\ref{gmrs13}) is fulfilled for all perturbations $\delta\overline{\omega}$ that conserve circulation and energy at first order, then according to Eq. (\ref{eqa4}), we know that inequality (\ref{gmrs8}) is fulfilled for all perturbations $\delta\rho$ that conserve circulation, energy, and normalization at first order; (ii) if there exists a perturbation $\delta\overline{\omega}_c$ that makes $\delta^{2}J[\delta\overline{\omega}]>0$, then the perturbation $\delta\rho_c$ given by Eq. (\ref{eqa3}) with $\delta\overline{\omega}=\delta\overline{\omega}_c$ makes $\delta^{2}J[\delta\rho]>0$. In conclusion, the stability criteria (\ref{gmrs8}) and (\ref{gmrs13}) are equivalent.

\section{The equation for the vorticity distribution} \label{sec_vd}

In the parametrization of Sec. \ref{sec_parc}, the mean vorticity $\overline{\omega}({\bf r},t)$ evolves according to Eqs. (\ref{intro8})-(\ref{intro9}) and the vorticity distribution $\rho({\bf r},\sigma,t)$ is then given by Eqs. (\ref{nge1})-(\ref{nge3}). It can be of interest to determine the relaxation equation satisfied by $\rho({\bf r},\sigma,t)$. According to Eq. (\ref{nge1}), we have
\begin{equation}
\ln\rho=-\sigma\Phi+\ln\chi(\sigma)-\ln\hat{\chi}(\Phi), \label{vd1}
\end{equation}
where $\Phi({\bf r},t)$ is related to $\overline{\omega}({\bf r},t)$ according to Eqs. (\ref{nge2})-(\ref{nge3}). Taking the derivative with respect to time and using Eq. (\ref{nge3}), we obtain
\begin{equation}
\frac{\partial\ln\rho}{\partial t}=-(\sigma-\overline{\omega})\frac{\partial\Phi}{\partial t}.  \label{vd2}
\end{equation}
Then, using Eq. (\ref{ge7}), we get
\begin{equation}
\frac{\partial\ln\rho}{\partial t}=(\sigma-\overline{\omega})C''(\overline{\omega})\frac{\partial\overline{\omega}}{\partial t}.  \label{vd3}
\end{equation}
Similarly, we have
\begin{equation}
\nabla\ln\rho=(\sigma-\overline{\omega})C''(\overline{\omega})\nabla\overline{\omega}.  \label{vd4}
\end{equation}
Combining these two relations, we obtain
\begin{equation}
{\partial \rho\over\partial t}+{\bf u}\cdot \nabla
\rho=\rho(\sigma-\overline{\omega})C''(\overline{\omega})\left ( {\partial \overline{\omega}\over\partial t}+{\bf u}\cdot \nabla
\overline{\omega}\right ).
\label{vd5}
\end{equation}
Substituting Eq. (\ref{intro8}) in Eq. (\ref{vd5}), we find that
\begin{eqnarray}
{\partial \rho\over\partial t}&+&{\bf u}\cdot \nabla
\rho\nonumber\\
&=&\rho(\sigma-\overline{\omega})C''(\overline{\omega})\nabla\cdot \biggl \lbrace D
\biggl\lbrack \nabla\overline{\omega}+{\beta(t)\over
C''(\overline{\omega})}\nabla\psi\biggr\rbrack\biggr\rbrace ,\qquad
\label{vd6}
\end{eqnarray}
Finally, using Eqs. (\ref{vd4}) and (\ref{ind}), the foregoing
equation can be rearranged in the form
\begin{eqnarray}
{\partial \rho\over\partial t}&+&{\bf u}\cdot \nabla
\rho\nonumber\\
&=&\frac{\rho(\sigma-\overline{\omega})}{\omega_2}\nabla\cdot \biggl \lbrace \frac{D\omega_2}{\rho(\sigma-\overline{\omega})}
\biggl\lbrack \nabla\rho+\beta(t)\rho (\sigma-\overline{\omega})\nabla\psi\biggr\rbrack\biggr\rbrace.\nonumber\\
\label{vd7}
\end{eqnarray}
Under this form, we see some analogies (but also crucial differences)
with the relaxation equation (\ref{mrs8}) related to the MRS
approach. Finally, in the case of a Gaussian prior, $\omega_2({\bf
r},t)=\Omega_2$ is constant and the foregoing equation reduces to
\begin{eqnarray}
{\partial \rho\over\partial t}&+&{\bf u}\cdot \nabla
\rho\nonumber\\
&=&\rho(\sigma-\overline{\omega})\nabla\cdot \biggl \lbrace \frac{D}{\rho(\sigma-\overline{\omega})}
\biggl\lbrack \nabla\rho+\beta(t)\rho (\sigma-\overline{\omega})\nabla\psi\biggr\rbrack\biggr\rbrace .\nonumber\\
\label{vd8}
\end{eqnarray}

\section{Derivation of the relaxation equation (\ref{two22new})}
\label{sec_der}

We write the dynamical equation as
\begin{eqnarray}
\frac{\partial\overline{\omega}}{\partial t}+{\bf u}\cdot \nabla\overline{\omega}=X, \label{ma1}
\end{eqnarray}
where $X$ is an  unknown quantity to be chosen so as to increase $S[\overline{\omega}]$ while conserving $E$ and $\Gamma$.
The time variations of $S$ are given by
\begin{equation}
\dot{S}=-\int C'(\overline{\omega}) X\, d{\bf r}. \label{ma2}
\end{equation}
On the other hand, the time variations of $E$ and $\Gamma$ are
\begin{eqnarray}
\dot{E}=\int X\psi \, d{\bf r}=0,\label{ma2b}\\
\dot{\Gamma}=\int X \, d{\bf r}=0. \label{ma3}
\end{eqnarray}

Following the Maximum Entropy Production Principle, we maximize $\dot{S}$ with $\dot{E}=\dot{\Gamma}=0$ and the additional constraint
\begin{equation}
\frac{X^2}{2}\le C({\bf r},t). \label{ma4}
\end{equation}
The variational principle can be written in the form
\begin{eqnarray} \label{ma5}
\delta \dot{S} - \beta(t) \delta \dot{E} - \alpha(t) \delta \dot{\Gamma} -
\int \frac{1}{D({\bf r},t)} \delta \left(\frac{X^2}{2} \right) d{\bf r}=0,\nonumber\\
\end{eqnarray}
and we obtain the following current
\begin{eqnarray}
X=-D \left(C'(\overline{\omega})+ \beta(t) {\psi} +\alpha(t) \right). \label{ma6}
\end{eqnarray}
Substituting Eq. (\ref{ma6}) in Eq. (\ref{ma1}), we obtain Eq. (\ref{two22new}).
 The Lagrange multipliers $\beta(t)$ and $\alpha(t)$ evolve so as to satisfy the constraints. Substituting Eq. (\ref{ma6}) in Eqs. (\ref{ma2b}) and (\ref{ma3}),  we obtain
the algebraic equations (\ref{two23new})-(\ref{two24new}). Finally, the H-theorem (\ref{hnm}) can be derived as follows. Substituting Eq. (\ref{ma6}) in Eq. (\ref{ma2}), we get
\begin{eqnarray}
\dot S=\int X \left (\frac{X}{D}+\beta(t)\psi+\alpha(t)\right )\, d{\bf r}. \label{ma7}
\end{eqnarray}
Then, using the constraints (\ref{ma2b})-(\ref{ma3}) we obtain
\begin{eqnarray}
\dot S=\int \frac{X^2}{D}\, d{\bf r}\ge 0. \label{ma8}
\end{eqnarray}

\end{document}